\documentclass[]{aastex631}

\usepackage{mhchem}
\usepackage{ulem}
\usepackage{threeparttable}

\shorttitle{MINDS: The JWST Mid-INfrared Disk Survey}
\shortauthors{MIRI/GTO disk (MINDS) team}
\graphicspath{{./}{figures/}}

\begin{document}

\title{MINDS: The JWST MIRI Mid-INfrared Disk Survey}


\author[0000-0012-3245-1234]{Thomas Henning}
\affiliation{Max-Planck-Institut f\"{u}r Astronomie (MPIA), K\"{o}nigstuhl 17, 69117 Heidelberg, Germany}
\altaffiliation{shared first authorship}

\author[0000-0012-3245-1234]{Inga Kamp}
\affiliation{Kapteyn Astronomical Institute, University of Groningen, 9700 AV Groningen, The Netherlands}
\altaffiliation{shared first authorship}

\author[0000-0001-9992-4067]{Matthias Samland}
\affil{Max-Planck-Institut f\"{u}r Astronomie (MPIA), K\"{o}nigstuhl 17, 69117 Heidelberg, Germany}

\author[0000-0001-8407-4020]{Aditya M. Arabhavi}
\affil{Kapteyn Astronomical Institute, University of Groningen, 9700 AV Groningen, The Netherlands}

\author[0000-0003-0386-2178]{Jayatee Kanwar}
\affil{Kapteyn Astronomical Institute, University of Groningen, 9700 AV Groningen, The Netherlands}
\affil{Space Research Institute, Austrian Academy of Sciences, Schmiedlstr. 6, A-8042, Graz, Austria}
\affil{TU Graz, Fakultät für Mathematik, Physik und Geodäsie, Petersgasse 16 8010 Graz, Austria}


\author[0000-0001-7591-1907]{Ewine F. van Dishoeck}
\affil{Leiden Observatory, Leiden University, 2300 RA Leiden, the Netherlands}
\affil{Max-Planck Institut f\"{u}r Extraterrestrische Physik (MPE), Giessenbachstr. 1, 85748, Garching, Germany}

\author[0000-0001-9818-0588]{Manuel G\"udel}
\affil{Dept. of Astrophysics, University of Vienna, T\"urkenschanzstr. 17, A-1180 Vienna, Austria}
\affil{ETH Z\"urich, Institute for Particle Physics and Astrophysics, Wolfgang-Pauli-Str. 27, 8093 Z\"urich, Switzerland}

\author{Pierre-Olivier Lagage}
\affil{Universit\'e Paris-Saclay, Universit\'e Paris Cit\'e, CEA, CNRS, AIM, F-91191 Gif-sur-Yvette, France}

\author{Christoffel Waelkens}
\affil{Institute of Astronomy, KU Leuven, Celestijnenlaan 200D, 3001 Leuven, Belgium}

\author{Alain Abergel}
\affil{Universit\'e Paris-Saclay, CNRS, Institut d’Astrophysique Spatiale, 91405, Orsay, France}

\author[0000-0002-4006-6237]{Olivier Absil}
\affil{STAR Institute, Universit\'e de Li\`ege, All\'ee du Six Ao\^ut 19c, 4000 Li\`ege, Belgium}

\author[0000-0002-5971-9242]{David Barrado}
\affil{Centro de Astrobiolog\'ia (CAB), CSIC-INTA, ESAC Campus, Camino Bajo del Castillo s/n, 28692 Villanueva de la Ca\~nada, Madrid, Spain}

\author{Anthony Boccaletti}
\affil{LESIA, Observatoire de Paris, Universit\'e PSL, CNRS, Sorbonne Universit\'e, Universit\'e de Paris, 5 place Jules Janssen, 92195 Meudon, France}

\author[0000-0003-4757-2500]{Jeroen Bouwman}
\affil{Max-Planck-Institut f\"{u}r Astronomie (MPIA), K\"{o}nigstuhl 17, 69117 Heidelberg, Germany}

\author[0000-0001-8876-6614]{Alessio Caratti o Garatti}
\affil{INAF – Osservatorio Astronomico di Capodimonte, Salita Moiariello 16, 80131 Napoli, Italy}
\affil{Dublin Institute for Advanced Studies, 31 Fitzwilliam Place, D02 XF86 Dublin, Ireland}

\author[0000-0003-2692-8926]{Vincent Geers}
\affil{UK Astronomy Technology Centre, Royal Observatory Edinburgh, Blackford Hill, Edinburgh EH9 3HJ, UK}

\author[0000-0001-9250-1547]{Adrian M. Glauser}
\affil{ETH Z\"urich, Institute for Particle Physics and Astrophysics, Wolfgang-Pauli-Str. 27, 8093 Z\"urich, Switzerland}

\author{Fred Lahuis}
\affil{SRON Netherlands Institute for Space Research, PO Box 800, 9700 AV, Groningen, The Netherlands}

\author[0000-0003-3217-5385]{Michael Mueller}
\affil{Kapteyn Astronomical Institute, Rijksuniversiteit Groningen, Postbus 800, 9700AV Groningen, The Netherlands}

\author{Cyrine Nehm\'e}
\affil{Universit\'e Paris-Saclay, Universit\'e Paris Cit\'e, CEA, CNRS, AIM, F-91191 Gif-sur-Yvette, France}

\author[0000-0003-3747-7120]{G\"oran Olofsson}
\affil{Department of Astronomy, Stockholm University, AlbaNova University Center, 10691 Stockholm, Sweden}

\author{Eric Pantin}
\affil{Universit\'e Paris-Saclay, Universit\'e Paris Cit\'e, CEA, CNRS, AIM, F-91191 Gif-sur-Yvette, France}

\author[0000-0002-2110-1068]{Tom P. Ray}
\affil{Dublin Institute for Advanced Studies, 31 Fitzwilliam Place, D02 XF86 Dublin, Ireland}

\author[0000-0003-4559-0721]{Silvia Scheithauer}
\affil{Max-Planck-Institut f\"{u}r Astronomie (MPIA), K\"{o}nigstuhl 17, 69117 Heidelberg, Germany}

\author[0000-0002-1368-3109]{Bart Vandenbussche}
\affil{Institute of Astronomy, KU Leuven, Celestijnenlaan 200D, 3001 Leuven, Belgium}

\author[0000-0002-5462-9387]{L. B. F. M. Waters}
\affil{Department of Astrophysics/IMAPP, Radboud University, PO Box 9010, 6500 GL Nijmegen, The Netherlands}
\affil{SRON Netherlands Institute for Space Research, Niels Bohrweg 4, NL-2333 CA Leiden, the Netherlands}

\author{Gillian Wright}
\affil{UK Astronomy Technology Centre, Royal Observatory Edinburgh, Blackford Hill, Edinburgh EH9 3HJ, UK}


\author[0000-0003-2820-1077]{Ioannis Argyriou}
\affil{Institute of Astronomy, KU Leuven, Celestijnenlaan 200D, 3001 Leuven, Belgium}

\author[0000-0002-0101-8814]{Valentin Christiaens}
\affil{Institute of Astronomy, KU Leuven, Celestijnenlaan 200D, 3001 Leuven, Belgium}
\affil{STAR Institute, Universit\'e de Li\`ege, All\'ee du Six Ao\^ut 19c, 4000 Li\`ege, Belgium}

\author{Riccardo Franceschi}
\affil{Max-Planck-Institut f\"{u}r Astronomie (MPIA), K\"{o}nigstuhl 17, 69117 Heidelberg, Germany}

\author[0000-0002-1257-7742]{Danny Gasman}
\affil{Institute of Astronomy, KU Leuven, Celestijnenlaan 200D, 3001 Leuven, Belgium}

\author[0000-0002-4022-4899]{Sierra L. Grant}
\affil{Max-Planck Institut f\"{u}r Extraterrestrische Physik (MPE), Giessenbachstr. 1, 85748, Garching, Germany}

\author{Rodrigo Guadarrama}
\affil{Dept. of Astrophysics, University of Vienna, T\"urkenschanzstr. 17, A-1180 Vienna, Austria}

\author{Hyerin Jang}
\affil{Department of Astrophysics/IMAPP, Radboud University, PO Box 9010, 6500 GL Nijmegen, The Netherlands}

\author[0000-0001-9526-9499]{Maria Morales-Calder\'on}
\affil{Centro de Astrobiolog\'ia (CAB), CSIC-INTA, ESAC Campus, Camino Bajo del Castillo s/n, 28692 Villanueva de la Ca\~nada,
Madrid, Spain}

\author[0000-0002-9385-9820]{Nicole Pawellek}
\affil{Dept. of Astrophysics, University of Vienna, T\"urkenschanzstr. 17, A-1180 Vienna, Austria}
\affil{Konkoly Observatory, Research Centre for Astronomy and Earth Sciences, E\"otv\"os Lor\'and Research Network (ELKH), Konkoly-Thege Mikl\'os \'ut 15-17, H-1121 Budapest, Hungary}

\author[0000-0002-8545-6175]{Giulia Perotti}
\affil{Max-Planck-Institut f\"{u}r Astronomie (MPIA), K\"{o}nigstuhl 17, 69117 Heidelberg, Germany}

\author[0000-0002-0100-1297]{Donna Rodgers-Lee}
\affil{Dublin Institute for Advanced Studies, 31 Fitzwilliam Place, D02 XF86 Dublin, Ireland}

\author{J\"urgen Schreiber}
\affil{Max-Planck-Institut f\"{u}r Astronomie (MPIA), K\"{o}nigstuhl 17, 69117 Heidelberg, Germany}
 
\author[0000-0002-6429-9457]{Kamber Schwarz}
\affil{Max-Planck-Institut f\"{u}r Astronomie (MPIA), K\"{o}nigstuhl 17, 69117 Heidelberg, Germany}

\author{Beno\^{i}t Tabone}
\affil{Universit\'e Paris-Saclay, CNRS, Institut d’Astrophysique Spatiale, 91405, Orsay, France}

\author[0000-0002-7935-7445]{Milou Temmink}
\affil{Leiden Observatory, Leiden University, 2300 RA Leiden, the Netherlands}

\author[0000-0002-3135-2477]{Marissa Vlasblom}
\affil{Leiden Observatory, Leiden University, 2300 RA Leiden, the Netherlands}


\author{Luis Colina}
\affil{Centro de Astrobiolog\'ia (CAB, CSIC-INTA), Carretera de Ajalvir, E-28850 Torrej\'on de Ardoz, Madrid, Spain}

\author{Thomas R. Greve}
\affil{DTU Space, Technical University of Denmark, Building 328, Elektrovej, 2800 Kgs. Lyngby, Denmark}

\author{G\"oran \"Ostlin}
\affil{Department of Astronomy, Oskar Klein Centre; Stockholm University; SE-106 91 Stockholm, Sweden}



\begin{abstract}

The study of protoplanetary disks has become increasingly important with the Kepler satellite finding that exoplanets are ubiquitous around stars in our galaxy and the discovery of enormous diversity in planetary system architectures and planet properties. High-resolution near-IR and ALMA images show strong evidence for ongoing planet formation in young disks. The JWST MIRI mid-INfrared Disk Survey (MINDS) aims to (1) investigate the chemical inventory in the terrestrial planet-forming zone across stellar spectral type, (2) follow the gas evolution into the disk dispersal stage, and (3) study the structure of protoplanetary and debris disks in the thermal mid-IR. The MINDS survey will thus build a bridge between the chemical inventory of disks and the properties of exoplanets. The survey comprises 52 targets (Herbig\,Ae stars, T\,Tauri stars, very low-mass stars and young debris disks). We primarily obtain MIRI/MRS spectra with high S/N ($\sim\!100-500$) covering the complete wavelength range from 4.9 to 27.9~$\mu$m. For a handful of selected targets we also obtain NIRSpec IFU high resolution spectroscopy (2.87-5.27~$\mu$m). We will search for signposts of planet formation in thermal emission of micron-sized dust – information complementary to near-IR scattered light emission from small dust grains and emission from large dust in the submillimeter wavelength domain. We will also study the spatial structure of disks in three key systems that have shown signposts for planet formation, TW\,Hya and HD\,169142 using the MIRI coronagraph at 15.5~$\mu$m and 10.65~$\mu$m respectively and PDS\,70 using NIRCam imaging in the $1.87~\mu$m narrow and the $4.8~\mu$m medium band filter. We provide here an overview of the MINDS survey and showcase the power of the new JWST mid-IR molecular spectroscopy with the TW\,Hya disk spectrum where we report the detection of the molecular ion \ce{CH3+} and the robust confirmation of \ce{HCO+} earlier detected with Spitzer. 

\end{abstract}



\section{Introduction} \label{sec:intro}

Over the last years, Kepler and now TESS have shown that the efficiency of planet formation has to be high, i.e., that every circumstellar disk around a young star is essentially forming planets. Recent near-IR imaging at 10~m class telescopes has revealed protoplanet candidates embedded in the birth environment of their disks \citep[e.g.,][]{Gratton2019, Hammond2023} and led to the confirmed detection of one system with two massive protoplanets inside a dust gap, PDS\,70 \citep{Keppler2018,  Mueller2018}. ALMA confirms the presence of a circumplanetary disk around PDS\,70\,c \citep{Benisty2021}. More indirectly, gas kinematics can indicate the potential presence of massive protoplanets in disks \citep{Pinte2018b, Teague2018, Pinte2023}. In addition, VLT/SPHERE, Gemini/GPI, Subaru and ALMA high-resolution images of planet forming disks show intriguing substructure that is often attributed to ongoing planet formation \citep[e.g., dust traps, rings, gaps and spirals –][]{Tamura2009,vanderMarel2013, ALMA2015, Benisty2015, vanBoekel2017, Andrews2018,Rich2022}. In young primordial disks, dust grains, gas and protoplanets interact with each other, leading to dust migration, trapping and thus producing this wealth of substructure \citep[][]{Andrews2020,Birnstiel2023}. 

In later stages of planet formation, both dust and gas are likely of secondary origin and could carry information on the composition of the planetesimals that are still contributing to the formation of terrestrial planets \citep{Wyatt2008}. Herschel and ALMA spectroscopy found traces of gas left in young debris disks \citep[e.g.,][]{Riviere2012, Roberge2013, Dent2014, Moor2017}; some of these systems are harboring cold dust in Kuiper belts next to warm dust in asteroid belts \citep{Matthews2014}.

The next challenge in planet formation studies is to provide a connection between the chemical composition of disks in the gas and solids and the composition of planets and their atmospheres. All of these recent findings underline the importance of characterizing the current chemical composition and structure of disks during the entire planet forming phase (well into the debris disk stage) and linking it to the planetary systems that are forming inside them \citep[e.g.,][]{Molliere2022}. This is the goal of the Guaranteed Time Observations (GTO) JWST MIRI mid-INfrared Disk Survey (MINDS, PI: Th.\ Henning), especially for the molecular inventory in the inner terrestrial planet-forming regions of disks. It bridges between the GTO programs that study embedded disks (JOYS, PI: E.\ van Dishoeck) and exoplanets (EXO, PI: P.-O.\ Lagage).

\subsection{Scientific goals of MINDS}
\label{sec:goals}

We will use JWST to (1) investigate the chemical inventory in the terrestrial planet-forming zone across stellar spectral type, (2) follow the gas evolution into the disk dispersal stage, and (3) study the structure of protoplanetary and debris disks in the thermal mid-IR. The MINDS survey will thus build a bridge between the chemical inventory of disks and the properties of exoplanets. In the subsequent paragraphs, we outline these in more detail.

\subsubsection{The chemical inventory} 
\label{sec:goal:chemicalcomposition}

The program aims at a comprehensive characterization of the dust and gas content of disks during the entire planet-forming phase and across stellar spectral types (A0-M8). After Spitzer and ground-based detections of H$_2$O, OH, HCN, C$_2$H$_2$ and CO$_2$ \citep[e.g.,][]{Carr2008, Salyk2008, Pascucci2009, Pontoppidan2010, Mandell2012, Pascucci2013}, we will search for less abundant/more complex molecules such as CH$_4$, C$_6$H$_6$, NH$_3$, HCO$^+$ or HNC and rare isotopologues (e.g., $^{13}$CO$_2$, H$_2^{18}$O, HDO) that better constrain the abundances of the major species. A full inventory of molecules in the main planet-forming region (inside a few au) provides important clues to the formation of even more complex molecules that eventually link back to the early history of our own Solar System, the origin and delivery of water on Earth, and, eventually, the emergence of life. In addition, we follow the evolution of detailed dust mineralogy enabled by the unprecedented high S/N in the spectra. Such a comprehensive inventory of dust properties and gas composition also helps to understand the compositional dichotomy of the inner disk between disks around very low-mass and solar-type stars first noted by \citet{Pascucci2009, Pascucci2013} based on Spitzer data.

\subsubsection{Disk dispersal} 

A series of atomic (neutral and ionized) and molecular lines as well as PAH and silicate features in the MIRI wavelength range will be used to determine the vertical structure of the disk surface layers and the presence of disk winds (e.g., [Ne\,{\sc ii}], [Fe\,{\sc ii}], [S\,{\sc i}], H$_2$). These tracers test different scenarios for inner disk evolution and clearing such as internal photoevaporation, winds, dust settling/growth and planetesimal formation \citep[e.g.,][]{Nomura2007, Lahuis2007, Pascucci2007, vanBoekel2009, BaldovinSaavedra2012, Sacco2012} as well as gas dispersal through accretion \citep[see e.g.,][]{Manara2023}. Indeed, the MIRI wavelength range covers many hydrogen recombination lines, which can be used to estimate mass accretion rates \citep[e.g., Franceschi et al.\ submitted,][]{Rigliaco2015}. 

We will also characterize the evolving gas composition well into the debris disk phase. The mid-IR traces ionized and atomic fine structure lines of sulphur, iron and molecular lines of SiO that are key in characterizing the nature of the gas and its link to planet formation \citep[e.g.,][]{Gorti2004}. The trace amounts of gas are often attributed to planetesimal collisions and/or collisions of terrestrial mass objects \citep[e.g.,][]{Lisse2009, Dent2014}. Depending on the gas temperature and the CO/H$_2$ ratio we may be able to discover molecular hydrogen emission from young debris disks.

\subsubsection{Disk structure in the mid-IR} 

VLT/NACO, VLT/SPHERE, GPI and ALMA images show frequently inner depleted regions in disks as well as substructures such as bright secondary rims, azimuthal asymmetries, and/or spiral density waves \citep[see][for recent PP\,{\sc vii} chapters]{BenistyPPVII2022,BaePPVII2022}. Such features could be linked to forming massive planets. MIRI imaging not only provides the missing link between the scattered light by small micron-sized grains and thermal emission of mm-sized grains, but also has the capability of detecting massive protoplanets \citep{Chen2022}. We will eventually investigate whether and how disk substructure and pebble transport are linked to the diversity in chemical composition (Sect.~\ref{sec:goal:chemicalcomposition}). 

\section{Sample selection} \label{sec:sample}

We selected our sample to contain young class\,II disks around stars of a wide range of spectral types in star-forming regions (Herbig\,Ae stars, T\,Tauri stars and very low-mass stars), and a sample of five young debris disks with CO sub-mm detections (e.g., 49\,Ceti, HD\,131835, HD\,21997). Besides a few individual key targets (e.g., TW\,Hya, HD\,169142, PDS\,70), we selected targets in the nearby (120-190~pc) Taurus, Ophiuchus, Chamaeleon and Lupus star-forming regions. After considering the MIRI/MRS brightness limits, the main selection criteria for our young class\,{\sc ii} disks were:
\begin{itemize}
    \item previous detection of molecular line emission and/or atomic fine structure lines with Spitzer \citep{Najita2003,Pascucci2009,Pontoppidan2010,Salyk2011,Pascucci2013}
    \item ground-based high spectral resolution CO ro-vibrational line profiles indicative of a disk origin \citep{Najita2003,Salyk2009, Brown2013,Garufi2014,vanderPlas2015}
    \item some spread over the various star-forming regions, but the primary focus is on Taurus
    \item existing (or scheduled) ALMA observations.
\end{itemize} 
Our sample covers a stellar mass range $0.1-3$~M$_\odot$ (see Table~\ref{tab:sample}): 5 Herbig disks (two of which close to edge-on), 33 T\,Tauri disks (four of which edge-on), 10 disks around very low-mass stars (VLMS\footnote{We use here a spectral type of M4 to distinguish between T Tauri and VLMS. However, this is not a well defined boundary and is used quite loosely in the literature.}, one close to edge-on, 2MASS-J04381486+2611399). 

For a few selected targets, we obtain additional NIRSpec IFU high-resolution spectroscopy: AA\,Tau (T\,Tauri), XX\,Cha (T\,Tauri), PDS\,70 (T\,Tauri, see below), HK \,Tau\,B (T\,Tauri, edge-on), PDS\,453 (T\,Tauri, edge-on), Flying Saucer (T\ Tauri, edge-on), HV\,Tau\,C (T\,Tauri, edge-on), and PDS\,144\,N (Herbig, edge-on). NIRSpec will provide additional information on key ice features and covers the main part of the CO ro-vibrational band. The observing strategy is detailed in Sect.~\ref{sec:obs-nirspec}. 

Our targets with special signposts of planet formation are TW\,Hya, PDS\,70, HD\,135344B, and HD\,169142. TW\,Hya is the closest known planet-forming disk at a distance of 60.14~pc \citep{Gaia2021} which shows rings in VLT/SPHERE and ALMA data down to au scales \citep{Andrews2016,vanBoekel2017}. It is thus a prime target to search for signposts of protoplanets in thermal emission (Sect.~\ref{sec:obs-corona}). Another key target in this respect is PDS\,70 with two gas giant protoplanets confirmed inside a dust gap \citep[$\sim\!17-54$~au,][Sect.~\ref{sec:obs-PDS70}]{Keppler2018, Mueller2018, Benisty2021}. HD\,135344B is a disk around an F4V-type star with a dust cavity of $\sim\!40$~au that shows intriguing rings and spiral structure in VLT/NACO and ALMA data \citep{Garufi2013,vanderMarel2016, Cazzoletti2018}. Last but not least, the well-investigated planet-forming disk around the star HD\,169142 is characterized by two gaps \citep[e.g.,][]{Quanz2013} and an inner disk ($\lesssim\!20$~au). Therefore, it shares some similarities with the PDS\,70 system. The disk is very well spatially and kinematically characterized by VLT/SPHERE and ALMA observations and an earlier claimed detection of a protoplanet at 37~au \citep{Gratton2019} is now confirmed by re-analysing VLT/SPHERE data \citep{Hammond2023}. 

\newpage
\startlongtable
\movetabledown=3cm
\begin{rotatetable}
\begin{deluxetable}{llllllrrrrlrl}
\tabletypesize{\tiny}
\tablecaption{Overview of the JWST MINDS disk sample.\label{tab:sample}}
\tablehead{ \colhead{Source} & \colhead{RA (ICRS)} & \colhead{DEC (ICRS)} & \colhead{Region} & \colhead{Disk}  & \colhead{SpT} & \colhead{$d$} & \colhead{$T_{\rm eff}$} & \colhead{$L_\ast$} & \colhead{$M_\ast$} & \colhead{$\log\,g$} & \colhead{[Fe/H]} & \colhead{ALMA/} \\
  \colhead{ }      & \colhead{ep$=$2015.5} & \colhead{ep$=$2015.5} &  \colhead{ }&  \colhead{ }&  \colhead{ }&  \colhead{[pc]} &  \colhead{[K]} &  \colhead{[L$_{\odot}$]} & \colhead{[M$_{\odot}$]} &  \colhead{[log(cm/s$^2$)]} &  \colhead{ }&  \colhead{SpitzerHR} }
\startdata
  49Cet$^{(2)}$			  & 01:34:37.88 & -15:40:35.0 & Argus Moving Group  & D & A1 & 58.26 & 8770 & 15.58 & 1.99 & 4.18 & -0.91 & AS\\
  CX-Tau$^{(1)}$    		  & 04:14:47.87 & +26:48:10.7 & Taurus & TT & M1.5 & 126.74* & 3487 & 0.34 & 0.33 &  & & AS \\
  CY-Tau$^{(1)}$ 		  & 04:17:33.74 & +28:20:46.4 & Taurus & TT & M1.5 & 124.35 & 3516 & 0.37 & 0.35 & 3.99 & 0.17 & AS\\
  BP-Tau$^{(1)}$ 		  & 04:19:15.84 & +29:06:26.5 & Taurus & TT & K5/7 & 128.28 & 3777 & 0.83 & 0.47 & 4.21 & -4.15 & AS\\
  FT-Tau$^{(1)}$ 		  & 04:23:39.20 & +24:56:13.9 & Taurus & TT & M2.8 & 129.96 & 3415 & 0.44 & 0.29 & 4.32 & -4.15 & AS\\
  DF-Tau$^{(1)}$ 		  & 04:27:02.80 & +25:42:22.0 & Taurus & TT & M3 & 176.45* & 3900 & 3.89 & 0.61 &  &  & AS\\
  DG-Tau$^{(1)}$ 		  & 04:27:04.70 & +26:06:15.7 & Taurus & TT & K6 & 130.21 & 4060 & 2.4 & 0.84 & 4.17 & -1.49 & AS\\
  HH30           		  & 04:31:37.47 & +18:12:24.5 & Taurus & TTe & K7 &  &  &  &  &  &  & A \\
  HK-Tau-B       		  & 04:31:50.58 & +24:24:16.4 & Taurus & TTe & M0.5 & 128.53* &  &  &  &  &  & AS \\
  DL-Tau$^{(1)}$ 		  & 04:33:39.09 & +25:20:37.8 & Taurus & TT & K7 & 159.53 & 4276 & 1.47 & 0.89 & 4.23 & -2.49 & AS\\
  DM-Tau$^{(1)}$ 		  & 04:33:48.75 & +18:10:09.7 & Taurus & TT & M2 & 144.8 & 3415 & 0.25 & 0.29 & 4.06 & -0.21 & AS\\
  AA-Tau$^{(1)}$   	       	  & 04:34:55.43 & +24:28:52.7 & Taurus & TT & K5 & 137.72 & 3762 & 0.72 & 0.47 & 4.63 & -3.91 & AS \\
  DN-Tau$^{(1)}$   	       	  & 04:35:27.38 & +24:14:58.6 & Taurus & TT & M1: & 127.29 & 3806 & 0.69 & 0.5 & 4.14 & -4.07 & AS\\
  2MASS-J04381486+2611399$^{(1)}$ & 04:38:14.89 & +26:11:39.6 & Taurus & VLMSe & M7 & 140.26* & 2840 & 0.003 & 0.06 &  & & AS \\
  HV-Tau-C       		  & 04:38:35.51 & +26:10:41.3 & Taurus & TTe & M1 &  &  &  &  &  &  & A \\
  2MASS-J04390163+2336029$^{(1)}$ & 04:39:01.64 & +23:36:02.6 & Taurus & VLMS & M6 & 125.89 & 3139 & 0.1 & 0.17 & 4.22 & 0.16 & AS\\
  2MASS-J04390396+2544264$^{(1)}$ & 04:39:03.97 & +25:44:26.0 & Taurus & VLMS & M7 & 140.97* & 2840 & 0.03 & 0.06 &  & & AS \\
  LkCa15$^{(1)}$ 		  & 04:39:17.80 & +22:21:03.1 & Taurus & TT & K5 & 154.83 & 4276 & 1.12 & 0.91 & 4.13 & -0.34 & AS\\
  DR-Tau$^{(1)}$ 		  & 04:47:06.22 & +16:58:42.6 & Taurus & TT & K5 & 186.98 & 4202 & 3.71 & 1.11 & 4.09 & -4.14 & AS\\
  HD32297           		  & 05:02:27.44 & +07:27:39.3 &  & D & A0 & 132.41 & 11250 & 24.42 &  & 4.5 & -0.66 & AS \\
  RW-Aur         		  & 05:07:49.57 & +30:24:04.9 & Taurus & TT & K1/5+K5 & 150* &  &  &  &  &  &  AS\\
  HD35929           		  & 05:27:42.79 & -08:19:38.5 &  & H & F2 & 380.36* &  &  &  &  &  & AS \\
  SY-Cha$^{(1)}$ 		  & 10:56:30.28 & -77:11:39.4 & Chamaeleon & TT & K5 & 180.78 & 4060 & 0.43 & 0.85 & 4.21 & -1.14 & AS\\
  TW-Hya$^{(1)}$ 		  & 11:01:51.82 & -34:42:17.2 & TW\,Hya Moving Group  & TT & K6 & 59.96 & 4000 & 0.34 & 0.75 & 4.05 & -0.51 & AS\\
  2MASS-J11071668-7735532 	  & 11:07:16.56 & -77:35:53.2 & Chamaleon & VLMS & M7.5/8 & 199.14 & 2664 & 0.04 &  & 3.78 & 0.79 &  \\
  2MASS-J11071860-7732516$^{(1)}$ & 11:07:18.46 & -77:32:51.6 & Chamaleon & VLMS & M5.5 & 197.52* & 3060 & 0.03 & 0.13 &  & & A \\
  2MASS-J11074245-7733593$^{(1)}$ & 11:07:42.34 & -77:33:59.4 & Chamaleon & VLMS &  &  & 3060 & 0.04 & 0.13 &  &  & A\\
  VW-Cha$^{(1)}$ 		  & 11:08:01.33 & -77:42:28.6 & Chamaeleon & TT & K7+M0 & 188.16* & 4060 & 1.62 & 0.75 &  & & AS \\
  2MASS-J11082650-7715550$^{(1)}$ & 11:08:26.37 & -77:15:55.1 & Chamaleon & VLMS & M5.75 & 195.78* & 3060 & 0.11 & 0.01 &  & & A \\
  2MASS-J11085090-7625135$^{(1)}$ & 11:08:50.81 & -76:25:13.7 & Chamaleon & VLMS & M4 & 192.24* & 3060 & 0.03 & 0.13 &  &  & A\\
  WX-Cha         		  & 11:09:58.57 & -77:37:09.0 & Chamaeleon & TT & M1+M5 &  &  &  &  &  &  & AS \\
  XX-Cha$^{(1)}$ 		  & 11:11:39.58 & -76:20:15.0 & Chamaeleon & TT & M3 & 194.64 & 3340 & 0.29 & 0.25 & 4.5 & -1.18 & AS\\
  Sz50           		  & 13:00:55.24 & -77:10:22.4 & Chamaeleon & TT & M3 &  &  &  &  &  &  & AS \\
  PDS70             		  & 14:08:10.11 & -41:23:52.9 & Upper Centaurus Lupus  & TT & K7 & 112.32 & 4138 & 0.38 &  & 4.15 & 0.11 & A \\
  HD131835       		  & 14:56:54.44 & -35:41:44.1 & Upper Centaurus Lupus & D & A2 & 130.28 & 8408 & 10.39 & 1.77 & 4.23 & -0.6 & A \\
  HD135344B$^{(2)}$ 		  & 15:15:48.42 & -37:09:16.4 & Upper Centaurus Lupus  & H & F8 & 135.41 & 6620 & 7.07 & 1.58 & 3.8 & -0.82 & AS\\
  HD138813       		  & 15:35:16.08 & -25:44:03.4 & Upper Scorpius  & D & A0 & 136.42 & 9900 & 26.88 & 2.23 & 4.14 & -1.42  & AS\\
  GW-Lup$^{(1)}$ 		  & 15:46:44.71 & -34:30:36.0 & Lupus & TT & M1.5 & 155.20* & 3632 & 0.33 & 0.41 &  & & AS \\
  PDS144N        		  & 15:49:15.55 & -26:00:49.6 &  & He & A2 & 130.00* &  &  &  &  &  & S \\
  IM-Lup$^{(1)}$ 		  & 15:56:09.19 & -37:56:06.5 & Lupus & TT & M0 & 153.81 & 4350 & 2.57 & 0.95 & 4.06 & -0.39 & AS\\
  2MASS-J15582981-2310077$^{(1)}$ & 15:58:29.80 & -23:10:08.1 & Upper Scorpius & TT & M3 & 136.81 & 3388 & 0.05 & 0.31 & 4.56 & 0.49 & AS\\
  2MASS-J16053215-1933159$^{(1)}$ & 16:05:32.14 & -19:33:16.3 & Upper Scorpius & VLMS & M5 & 158.17 & 3090 & 0.03 & 0.13 & 3.91 & -0.5 & AS\\
  Sz98$^{(1)}$      		  & 16:08:22.48 & -39:04:46.8 & Lupus & TT & M0.4 & 156.27* & 4060 & 1.51 & 0.67 &  &  & A\\
  V1094Sco$^{(1)}$  		  & 16:08:36.16 & -39:23:02.8 & Lupus & TT & K6/K5 & 152.44 & 4205 & 1.15 & 0.83 & 4.17 & -0.7 & A\\
  SR21           		  & 16:27:10.27 & -24:19:13.1 & Ophiucus & TT & G1 & 136.43* &  &  &  &  &  & AS \\
  IRS46          		  & 16:27:29.42 & -24:39:16.6 & Ophiucus & TT &  &  &  &  &  &  &  & AS \\
  2MASS-J16281370-2431391 	  & 16:28:13.75 & -24:31:40.0 & Ophiucus & TTe &  &  &  &  &  &  & & A  \\
  RNO90          		  & 16:34:09.16 & -15:48:17.2 & Ophiucus & TT & G5 & 114.96* &  &  &  &  & & AS  \\
  WA-Oph6        		  & 16:48:45.62 & -14:16:36.2 & Ophiucus & TT &  & 122.53* &  &  &  &  &  & AS \\
  PDS453         		  & 17:20:56.12 & -26:03:31.0 &  & He & F2 & 129.05* &  &  &  &  &  &  \\
  HD169142          		  & 18:24:29.78 & -29:46:49.9 &  & H & F1 & 115.36 & 7209 & 6.21 & 1.51 & 4.12 & -1.88 & AS\\
  HD172555       		  & 18:45:26.98 & -64:52:18.9 & $\beta$~Pic Moving Group  & D & A7 & 28.78* &  &  &  &  &  & AS \\
\enddata
\tablecomments{Coordinates are taken from Gaia DR2, effective temperature $T_{\rm eff}$, stellar luminosity $L_\ast$ and stellar mass $M_\ast$ come either from (1) Testi et al. (2022), (2) Kaeufer et al. (2023), or Gaia DR3; distance $d$, $\log\,g$ and [Fe/H] are retrieved from Gaia DR3. Spectral types (SpT) are collected from Simbad. The last column denotes the existence of either ALMA observations (A), Spitzer-IRS High resolution data (S), or both (AS). Disk types are defined as follows: T Tauri disk (TT); Debris disk (D); Herbig disk (H); very low-mass star disk (VLMS); edge-on disk (extra letter `e' appended). A `*' symbol in the distance column means that it has been estimated from the parallax.}
\end{deluxetable}
\end{rotatetable}
\newpage

\section{Observing strategy} 
\label{sec:obs-strategy}

We detail here our observing strategy for the MIRI \citep[][]{Wright2015, Wells2015, Labiano2021, Wright2023}, NIRSpec \citep{Jakobsen2022} and NIRCam \citep{Rieke2005, Burriesci2005}  observations of our MINDS survey. At this stage, all sources listed in Table~\ref{tab:sample} have been observed (MIRI/MRS, NIRSpec/IFU, NIRCam imaging and MIRI coronagraphy) except for the MIRI/MRS observation of SR\,21 and J04390396+2544264, the NIRSpec observation of the flying saucer and the MIRI/MRS and coronagraphic observations of HD\,163296. The published data products will be made available to the community via our website https://minds.cab.inta-csic.es.

\subsection{MIRI/MRS spectroscopy}
\label{sec:obs-miri}

We use MIRI/MRS ($R\!\sim\!3000$) and cover the entire wavelength range (all four channels) with S/N values of $100-500$ on the continuum. We apply a 4-point point-source dither pattern and typical exposure times between 1800 and 3600~sec on source. For a few bright targets (DG\,Tau\,B, RNO\,90, SR\,21, PDS\,144\,N, HD\,135344B), we use a non-standard choice of 3 groups to minimize saturation primarily in the short wavelength channels.

For sources that were identified as extended (DG\,Tau\,B, Flying Saucer, HH\,30, HD\,135344\,B, TW\,Hya), we implement a dedicated background at a clean offset position. The integration time is equal to that in a single dither position. In those cases, we also use the 4-point extended-source dither pattern. We also added this for the two sources WX\,Cha and XX\,Cha which were scheduled at the beginning of JWST observations, so that we could assess the importance of a dedicated background for the quality of the data reduction. For other sources, we extract the background from the on-source IFU image (see Sect.~\ref{sec:background}).

Most of our targets are bright and we decided to go without a dedicated target acquisition (TA). For a few key sources where we expect very faint emission and/or search for disk substructures, we implemented TA using a neutral density filter (FND) and fast readout (FASTR1); MIRI/MRS observations of two targets, PDS\,70 and HD\,169142 are described in more detail below.

Since many of our sources belonging to the same star-forming region can be grouped, we benefit from shorter slew times, thus reducing the overheads (Smart Accounting).

\subsection{NIRSpec/IFU spectroscopy}
\label{sec:obs-nirspec}

We take high-resolution ($R\!\sim\! 2700$) spectra in the long wavelength channel of NIRSpec, G395H/L290LP, $2.87–5.27~\mu$m for six sources: AA\,Tau, HK\,Tau\,B, PDS\,453, XX\,Cha, PDS\,70, and the Flying Saucer. Again, we skip TA and we use a 4-point-nod dither. With $600-700$~sec on source, we reach a S/N of a $100-200$ for a 100~mJy (continuum) source. 

\subsection{MIRI coronagraphy of TW\,Hya}
\label{sec:obs-corona}

Detailed coronagraphic simulations based on recent SPHERE images show that the best wavelength for detecting disk substructure is 15.5 $\mu$m (E.\ Pantin, private communication). This wavelength is also better suited for giant planet detection because it avoids the strong silicate band emission originating from dust in the inner disk. In 1800~sec, we reach a sensitivity of $\sim\,5\,\mu$Jy/10-$\sigma$. We take two position angles ($10-14^{\rm o}$ difference) and also use a comparison star for PSF subtraction, 26\,Crt. 
Calibration background sequences have been added in order to suppress the thermal background straylight discovered during MIRI commissioning \citep{Boccaletti2022}.

\subsection{A pan-instrument view of PDS70}
\label{sec:obs-PDS70}

We decided to study this iconic system hosting two gas giant protoplanets with three different instruments onboard JWST to characterize (i) the warm gas and dust composition/sizes in the inner disk (MIRI/NIRSpec IFU spectroscopy), (ii) the protoplanet properties (NIRCam direct imaging) and to (iii) try to detect the circumplanetary disks around the protoplanets (MIRI/NIRCam).

MIRI and NIRSpec IFU spectroscopy are implemented in the same way as for the other objects in our sample. The main difference is that we do TA with MIRI to ensure an optimum fringe correction, and that we integrate deeper, $\sim\!1$~hr on source ($\sim\,63$~min with MIRI and $\sim\,77$~min with NIRSpec). The NIRSpec data will give us access to the full CO ro-vibrational band (contrary to MIRI).

NIRCam imaging is performed in two filters, F187N and F480M. To avoid saturation in direct imaging, we use the subarray SUB64P (64x64 pixels corresponding to 2\arcsec\,x\,2\arcsec) to allow fast readout (RAPID). We use two position angles (PA1 and PA2),  separated by 5$^{\rm o}$ (the maximum possible to avoid scheduling issues) and for each of them 142 integrations each with 7 groups, resulting in a total exposure time of $\sim\!5$~min per PA. The primary filter F187N is chosen to detect the protoplanets thermal emission; it also covers Pa\,$\alpha$, a tracer of gas accretion onto a forming gas giant planet \citep{Aoyama2018}. We aim to use the `free' extra filter F480M to search for CO ro-vibrational emission, which could be due to the presence of a circumplanetary disk \citep{Oberg2023}.

\subsection{Characterizing the HD\,169142 inner disk and protoplanet}
\label{sec:obs-HD169142}

We will obtain MIRI/MRS data for this source in the same way as for our T\,Tauri disks. The difference is that we will obtain in a single observing sequence both coronagraphic imaging and MIRI/MRS observations (PID\,4525). This choice is motivated to image and potentially spectrally characterize the CPD of the protoplanet, in addition to constraining the chemistry of the inner disk. The coronagraphic observation is a 36~min integration with the 4-quadrant phase-mask at $10.65~\mu$m, using fast readout mode and a 9-point small grid dither; we use a PSF reference star and a dedicated background observation as recommended in \citet{Boccaletti2022}. For MIRI/MRS observations, we use a short 5~min exposure to avoid saturation and a 4-point extended-source dither and dedicated TA using the F1500W filter. This data can be used to study spatially extended mid-IR emission as well as the mid-IR spectrum.

\subsection{A serendipitous search for asteroids}
\label{sec:obs-asteroid}

MIRI/MRS observations allow for parallel imaging at no extra cost (except the data volume rate), so-called `simultaneous imaging'. Since our targets are all close to the ecliptic, we decided to implement a parallel imaging program focusing on the search for asteroids. Based on the expected typical temperatures, we adopted the F1280W filter for that serendipitous science goal. The integrations and groups for the imager are adjusted so that the total integration time is smaller but as close as possible to the main MIRI/MRS observations. The field of view of the parallel images is 74\arcsec\,x\,113\arcsec. We implement this for all our targets except in cases where we exceed the allowed data volume (data  excess $>\,15\,000$~MB). 

\section{Data reduction} \label{sec:datareduction}

We describe in this section the global procedure adopted for the reduction of our MIRI/MRS MINDS data, highlighting the choices made for the TW\,Hya dataset. For the reduction of our data acquired with a different JWST instrument or different MIRI observing mode (i.e., NIRSpec/IFU, MIRI coronagraphy and NIRCAM imaging), we refer the interested reader to our upcoming papers presenting the respective datasets.

\subsection{Standard pipeline}

The MIRI/MRS MINDS data are processed from the uncalibrated raw data files using the JWST pipeline \citep[version 1.12.5,][for TW\,Hya]{Bushouse2022} and Calibration Reference Data System (CRDS) context (\texttt{jwst\_1146.pmap} for TW\,Hya). A stray light correction is applied with the standard pipeline to remove contamination produced by internal reflections within the MIRI detector arrays.
Our reduction procedure was complemented with routines from the VIP package \citep{GomezGonzalez2017,Christiaens2023}, used in particular for enhanced bad pixel correction (performed after stage~2, yielding less spikes in the final spectra), and source centroid identification for faint targets (Sect.~\ref{sec:extraction}).

\subsection{Background subtraction}
\label{sec:background}

To remove the background in the absence of dedicated background observations, we considered three options, the first two operate on the rate-files at the detector stage (i.e., as for the subtraction of dedicated background images), the last on the built spectral cubes: (a) carry out a direct pair-wise dither subtraction, (b) leverage the four-point dither pattern to obtain a first guess on the background map, then refine it using a median-filter which both smoothes the background estimate and removes residual star signals from it, or (c) estimate the background in an annulus directly surrounding the aperture used for photometry extraction.
In general, the annulus background subtraction was the most robust and provides consistently good results without introducing extra noise. For faint targets it is worth trying a combination of (b) with a small aperture to minimize the impact of artefacts introduced by using the dithers for building a subtraction template. For datasets where dedicated background observations are available, such as TW\,Hya, we also tried to subtract the dedicated background frames on the detector level. However, due to their limited exposure time compared to the science sequence, they introduce more noise than when performing an annulus subtraction in the extracted image cube. Hence, spectra shown in this work make use of the annulus subtraction approach.

\subsection{Fringe correction}

We apply three fringe corrections to the data: the standard fringe flat correction (\texttt{flat\_field}) that takes place in stage~2 of the pipeline, a residual fringe correction (\texttt{residual\_fringe\_step}) between stages~2 and 3 \citep{Argyriou2020}, and a spectrum-level fringe correction after stage~3 (Kavanagh et al. in prep). The final spectrum-level fringe correction is done on individual sub-bands, extracted from stage~3, to avoid any residuals that may be produced by stitching sub-bands together. For sources with TA, we perform fringe removal using the calibration files provided by \citet{Gasman2023}. Since most of our targets are taken without TA, using the asteroid observations for fringe correction as done in \citet{Pontoppidan2023} is not straightforward.

\subsection{Spectrum extraction}
\label{sec:extraction}

The MIRI Medium Resolution Spectroscopy (MRS) mode is characterized by four channels, covering the 4.9$-$27.9~$\mu$m spectral range. Each channel has three sub-bands, SHORT (A), MEDIUM (B) and LONG (C) resulting in a full spectrum composed by twelve sub-bands \citep{Wells2015}. The spectral resolving power changes from $R\!=\!3100-3750$ (channel 1, $4.9-7.65~\mu$m) to $R\!=\!1330-1930$ (channel 4, $17.7-27.9~\mu$m). The spectrum is extracted on source using an aperture of 2~FWHM (2 times $1.22 \times \lambda/D$) by default, with the centroid of the source identified through a Gaussian fit in a weighted average image for each sub-band. This is different from \citep{Pontoppidan2023}, who use 2.8~FWHM. Generally, the wavelength calibration is accurate to better than $\sim$1 spectral resolution element ($\sim$10-30~km~s$^{-1}$; \citealt{Argyriou2023}). 

\section{Modelling approach} 
\label{sec:models}

For the qualitative and quantitative analysis and interpretation of the JWST data, we use a range of models with different levels of complexity, ranging from 0D slab models to full 2D radiation thermo-chemical disk models \citep[for details see also][]{Kamp2023}. Fast retrieval for molecular emission features is done with 0D slab models, while grids or series of 2D radiative transfer and/or thermo-chemical models allow us to assess the impact of certain physical parameters on the disk spectra and images in a forward modeling manner. Below, we describe the various models in more detail.

\subsection{Slab models}

For the fast retrieval, we use 0D slab models calculated with either the RADEX code \citep{vanderTak2007} or the {\sc ProDiMo} code \citep{Thi2013, Woitke2018}. Slab models are generally run for each molecular species separately. The slab models assume a homogeneous medium with a constant gas temperature $T_{\rm gas}$ and species density. Multiple collision partners (e.g., H, \ce{H2}, He, \ce{e-}) and their corresponding densities $n_{\rm c}$ can be specified. In addition, we can include IR pumping, for example from thermal dust emission. The level populations are solved either in LTE \citep[using pre-tabulated partition functions from HITRAN2020,][]{Gordon2022} or non-LTE \citep{Song2015, Ramirez-Tannus2023}. The models use a simple 1D radiative transfer (plane-parallel slab geometry) to calculate the emergent intensity from a slab with a species column density $N_{\rm sp}$. For very high column densities, line-overlap between lines of a specific molecule can become important. We treat this assuming a combination of thermal and turbulent broadening \citep[e.g.,][]{Tabone2023}.

An overview of the molecular data used for calculating the mid-IR spectra is shown in Table~\ref{tab:molecular-data}. In some cases, collision cross sections exist and have been compiled, e.g., for CO, \ce{CO2}, OH, and HCN. In those cases non-LTE effects have been investigated and quantified; the respective references listed in the table contain more detailed information.

Such 0D slab models are computationally inexpensive and we can run large grids for each molecule and isotopologue. For the retrieval part of our project, we use a refined $\chi^2$ minimalization approach. Grids of selected molecular 0D slab models as well as the modeling software (python) are made available to the community in Arabhavi et al.\ (submitted). In order to provide an optimal result, the fitting procedure has to be constrained to relatively small wavelength ranges per molecule; ideally those ranges are free of emission from other molecules. Since this is often hard to find, we also use an iterative approach. We first fit the molecule being the strongest contributor to the line emission, then we subtract that best fit spectrum and proceed to the next molecule. In this way, we can successively proceed from the strongest to the weakest emission \citep[see e.g.,][]{Grant2023, Schwarz2023}.

For specific molecules which have lines distributed over the full MIRI wavelength range, it can be impossible to fit the emission at all wavelengths with a single set of 0D slab model parameters. This has been noted already for water in Spitzer spectra \citep{Liu2019, James2022}, but was hard to quantify due to the strong blending of lines at a resolution of $R\,\sim\,600$. With the increased spectral resolution of MIRI, we find now clear evidence for radial temperature gradients from fitting 0D slab models to a series of wavelength intervals \citep[e.g.,][]{Gasman2023}.

\addtocounter{table}{1}
\begin{table}[]
\caption{Molecules with emission in the mid-IR spectral region used in the 0D slab models and thermo-chemical disk models. Molecules for which we have isotopologue data are denoted by a superscript $^{(1)}$. For LTE treatment, we only list the number of lines. $^\dagger$ A sub-selection has been made for up to $v=5$ and $J=60$ in each vibrational level of the ground electronic state. $^\ast$ A sub-selection has been made for the two lowest vibrational levels, $v=0$ and $v=1$.}    \centering
    \begin{tabular}{lllll}
    \hline
    \hline
  molecule  & treatment & \# levels & \# lines & reference\\
    \hline
\ce{o-H2O} & non-LTE & 411 & 7597 & LAMDA\\
\ce{p-H2O} & non-LTE & 413 & 7341 & LAMDA\\
\ce{OH} & non-LTE & 412 & 2360 & \citet{Tabone2021}$^{\ast}$\\
\ce{OH}$^{(1)}$ & LTE &  & 56905 & HITRAN2020\\
\ce{O2}$^{(1)}$ & LTE &  & 29793 & HITRAN2020\\
\ce{CO} & non-LTE & 300 & 1470 & \citet{Thi2013, Song2015}$^{\dagger}$\\
\ce{CO2} & non-LTE & 640 & 3698 & \citet{Bosman2017}\\
$^{13}$\ce{CO2} & non-LTE & 640 & 3387 & \citet{Bosman2017}\\
\ce{CO2}$^{(1)}$ & LTE &  & 539603 & HITRAN2020\\
\ce{H2CO}$^{(1)}$ & LTE &  & 44601 & HITRAN2020\\
\ce{CH4}$^{(1)}$ & LTE &  & 445749 & HITRAN2020\\
\ce{C2H2}$^{(1)}$ & LTE &  & 83967 & HITRAN2020\\
\ce{C2H4}$^{(1)}$ & LTE &  & 77631 & HITRAN2020\\
\ce{C2H6}$^{(1)}$ & LTE &  & 70623 & HITRAN2020\\
\ce{C3H4} & LTE &  & 9906 & Arabhavi et al., submitted\\
\ce{C4H2} & LTE &  & 251245 & HITRAN2020\\
\ce{C6H6} & LTE &  & 54608 & Arabhavi et al., submitted\\
\ce{CH3OH} & LTE &  & 19897 & HITRAN2020\\
\ce{NH3}$^{(1)}$ & LTE &  & 90396 & HITRAN2020\\
\ce{NO}$^{(1)}$ & LTE &  & 384305 & HITRAN2020\\
\ce{N2O}$^{(1)}$ & LTE &  & 160478 & HITRAN2020\\
\ce{HCN} & non-LTE & 602 & 4622 & \citet{Bruderer2015}\\
\ce{HCN}$^{(1)}$ & LTE &  & 151484 & HITRAN2020\\
\ce{HC3N} & LTE  &  & 248273 & HITRAN2020\\
\ce{CH3CN} & LTE &  & 3572 & HITRAN2020\\
\ce{H2S}$^{(1)}$ & LTE &  & 54228 & HITRAN2020\\
\ce{SO}$^{(1)}$ & LTE &  & 44264 & HITRAN2020\\
\ce{SO2}$^{(1)}$ & LTE &  & 975058
 & HITRAN2020\\
\ce{SO3} & LTE &  & 14295 & HITRAN2020\\
\ce{CS}$^{(1)}$ & LTE &  & 2078 & HITRAN2020\\
\ce{OCS}$^{(1)}$ & LTE &  & 37461 & HITRAN2020\\
\ce{CH3+} & LTE &  & 16200 & \citet{Changala2023}\\
\hline
    \end{tabular}
    \label{tab:molecular-data}
\end{table}

\begin{figure}[h]
    \centering
    \includegraphics[width=\textwidth]{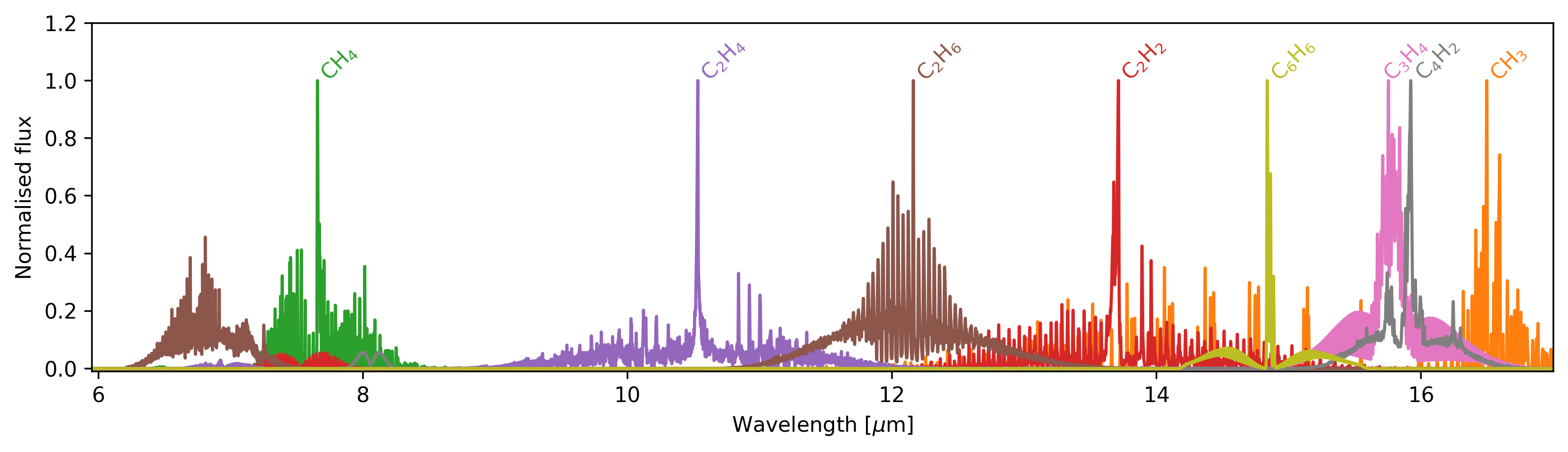}\\[-3mm]
    \caption{{\sc ProDiMo} LTE slab models with a series of hydrocarbons from the HITRAN2020 database \citep{Gordon2022} and from Arabhavi et al., submitted; line strengths have been normalized to the strongest intensity.}
    \label{fig:hydrocarbon-slab}
\end{figure}

Given the higher sensitivity and spectral resolution of JWST/MIRI, we expect to see higher excitation weaker bands (mostly Q-branches of overtones or combination bands) as well as isotopologues \citep[see for the example of \ce{CO2},][]{Bosman2018, Bosman2022b}. The mid-IR is also very rich in more complex hydrocarbons that could have lower abundances than \ce{C2H2}, but still be detectable mainly through their pronounced Q-branches (Fig.~\ref{fig:hydrocarbon-slab}).

\subsection{Simple two-component disk models}

The molecular emission, continuum and dust solid-state features are sometimes difficult to disentangle. In those cases, we employ simplified disk models, that account for realistic dust opacities, radial temperature gradients and capture the two-layer nature of massive planet-forming disks. Examples hereof are dust retrieval tools following the approach of \citet{Juhasz2009} and the CLIcK tool for fitting spectra in the presence of quasi-continua from blended line emission on top of dust continua \citep{Liu2019}.

The two-layer disk models calculate the continuum emission from annuli in an optically thick midplane emitting as a black body with a radial temperature gradient and an optically thin atmosphere (surface) which gives rise to the solid-state emission bands. The optically thick part can also contain a puffed up inner rim component. Molecular emission has been included in such two-layer models by assuming that the surface layer itself is optically thin in the continuum, i.e., the molecular emission at mid-IR wavelengths is not attenuated by dust.

\begin{figure}[h]
    \centering
    \includegraphics[width=0.5\textwidth]{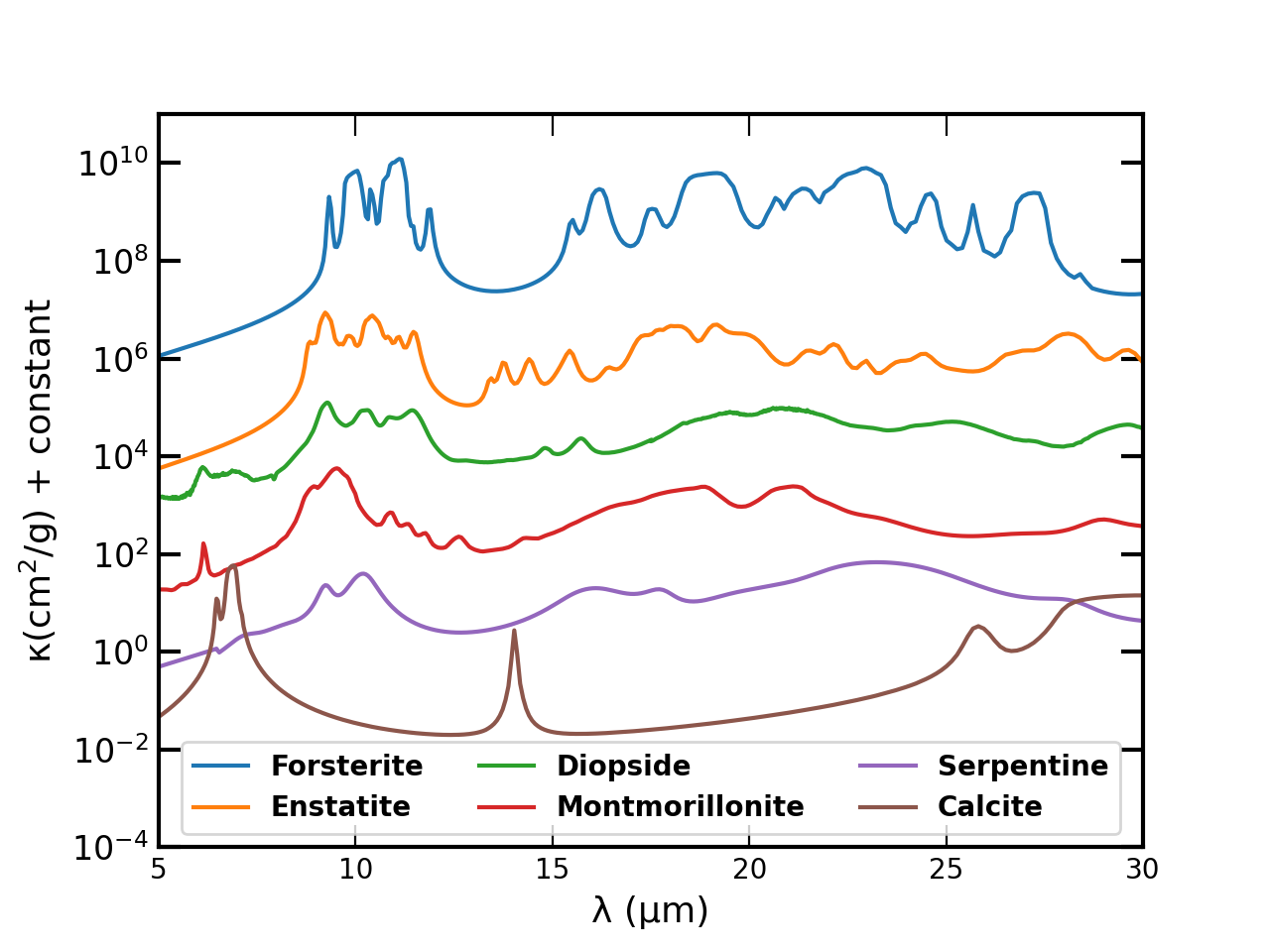}\\[-2mm]
    \caption{Opacities of various key minerals showing the richness of the MIRI wavelength range, especially the potential to detect rare minerals such as diopside, montmorillonite, serpentine and calcite.}
    \label{fig:dust-opacities}
\end{figure}

We evaluate the opacities of a number of minerals using DHS \citep{Min2005} and a transmission absorption spectrum for diopside. The optical constants are taken from \citet[][enstatite]{Zeidler2015}, \citet[][forsterite]{Suto2006}, \citet[][hydrosilicates]{Koike1990}, \cite[][diopside, see also \citealt{Posch2007}]{Koike2000}. Figure~\ref{fig:dust-opacities} shows that high S/N spectra in the MIRI wavelength range give unique access to rare dust species such as hydrous minerals (montmorillionite, serpentine) and calcite. These are minor constituents of meteorites in the solar system \citep{Suttle2021}.

\subsection{2D/3D disk models}

The SEDs of optically thin debris disks are modelled using the {\sc SONATA} code \citep{Pawellek2015}. For modeling debris disk images (e.g., MIRI IFU data) the 3D code MODERATO is used \citep{Pawellek2024}. Optically thick planet-forming disks are modelled using Monte-Carlo radiative transfer codes such as RADMC-3D \citep{Dullemond2012}, 
MCMAX-3D \citep{Min2009}, or 2D thermo-chemical disk codes such as DALI \citep{Bruderer2012} and {\sc ProDiMo} \citep{Woitke2009, Kamp2010, Thi2011b, Woitke2016}.

Since the focus of the MINDS program is the gas emission, we describe here briefly the power and drawbacks of thermo-chemical disk codes in some more detail. For details on the code structure, specific features and typical model set-ups, we refer to the respective papers of DALI and {\sc ProDiMo} above.

The typical runtime of a full thermo-chemical model is of the order of a few hours to a day (depending on the grid resolution, size of chemical network). This makes it immediately clear that such models cannot be used in straightforward Bayesian retrievals. In a fully consistent thermo-chemical model, the dust and gas structure are coupled through the radiative transfer and the chemistry; hence the abundances of atoms, ions and molecular species are constrained by the disk structure and assumptions on the dust properties. The same is true for the gas temperature, which is a result of the heating/cooling balance (which depends on species abundances). Dust settling and its treatment plays a key role as it determines the degree to which large grains settle and hence the dust opacity in the surface layers that we can observe with MIRI. DALI and {\sc ProDiMo} use different approaches. DALI uses typically two grain components, a small and a large one, where the latter is settled to a lower scale height (free parameter). {\sc ProDiMo} allows also for a parametrized settling, but most often uses the settling model by Dubrulle \citep[][see also \citealt{Schraepler2004}]{Dubrulle1995} or Riols settling \citep{Riols2018}. The former settles the dust grains according to the midplane sound speed, while the latter uses the local sound speed \citep{Woitke2023}. In the past, ad-hoc assumptions on molecular abundances, gas-to-dust ratios and/or gas temperatures have been employed to match mid-IR molecular emission spectra from disks. However, a wealth of studies over the past decade has improved our understanding of how the mid-IR molecular emission depends on inner disk structure, dust evolution, molecular shielding, dust opacities, gas temperatures, and element abundances \citep[e.g.,][]{Bethell2009,  Antonellini2015, Woitke2018, Greenwood2019, Woitke2019, Anderson2021, Bosman2022a, Bosman2022b, Kamp2023}. Taking into account as much of these improvement as possible, \citet{Woitke2023} were able to capture the key features of the MIRI spectrum of EX\,Lup \citep{Kospal2023} with a thermo-chemical disk model using {\sc ProDiMo}. 

Due to the shear number of lines and line density in the near- and mid-IR, an important component for predicting JWST spectra is a fast line radiative transfer. This led to the development of standalone codes such as RADLite \citep{Pontoppidan2009} and FLiTs \citep{Woitke2018}, which generally take the density, abundance and temperature structures from an external 2D disk code. Both line radiative transfer codes take into account the large velocity gradients across grid cells in the inner disk as well as the dust opacity. FLiTs also accounts consistently for opacity overlap between lines; this can be important for the Q-branches of species such as \ce{CO2}, HCN, \ce{C2H2} (also with their respective isotopologues) and OH/water \citep{Pontoppidan2009, Woitke2018}. Within the thermo-chemical codes, modules have been added to allow approximative calculation of realistic mid-IR emission spectra \citep[e.g., for DALI,][]{Bosman2017}.

\begin{figure}
    \centering
    \includegraphics[width=\textwidth]{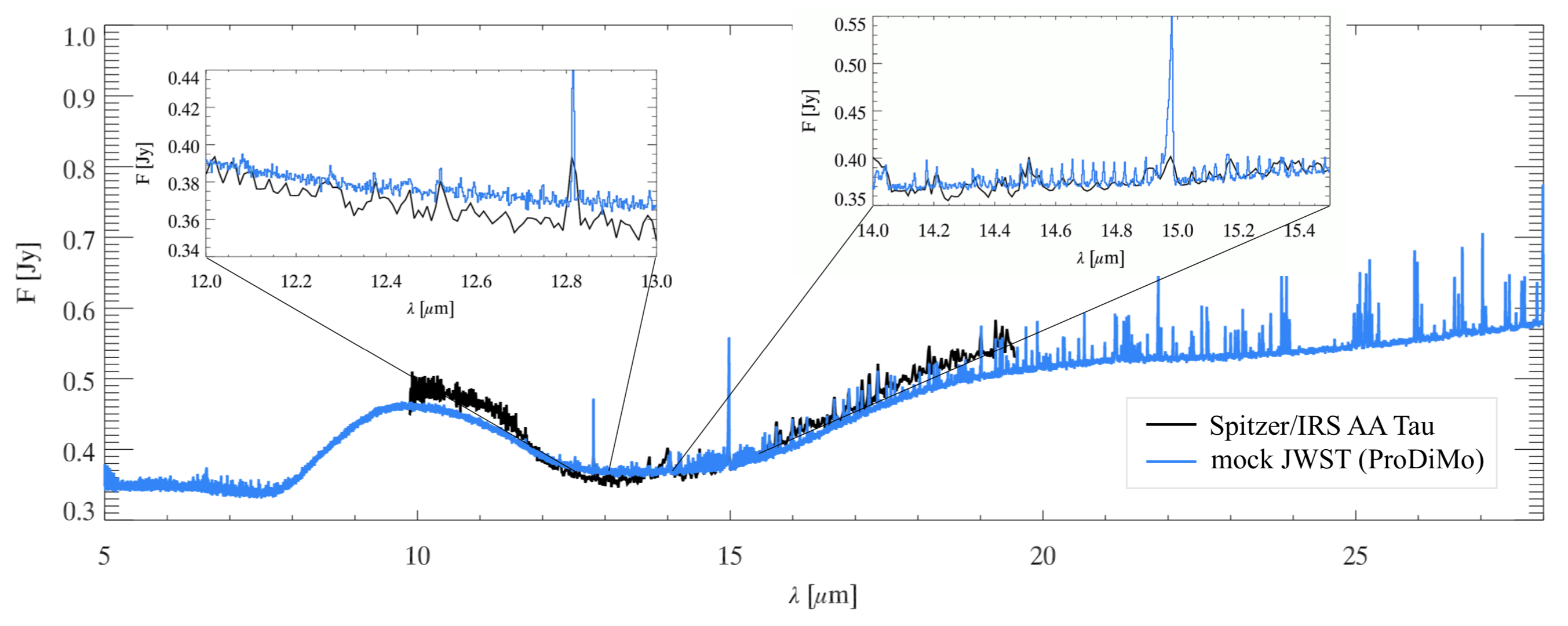}\\[-2mm]
    \caption{Simulated mid-IR spectrum ($R\!=\!3000$) for the disk around the star AA\,Tau including the full suite of atomic, ionic and molecular cooling lines (blue, DIANA SED model, S/N=300, $\sigma\!=\!2$~mJy). The Spitzer high resolution IRS spectrum ($R\!\sim\!600$) is shown in black (shifted by +50~mJy). The two insets cover the region of \ce{H2O} and [Ne\,{\sc ii}] at 12.8~$\mu$m and the \ce{CO2} region at 15~$\mu$m.}
    \label{fig:AATau_simulation}
\end{figure}

Such forward models include the complex interplay between gas and dust in realistic disk geometries benchmarked against a large range of multi-wavelength observations \citep[e.g.,][]{Woitke2019}. It is hence tempting to use these models to predict where the new discovery space for JWST is. In this way, \citet{Bosman2017} showed the diagnostic power of detecting the rarer isotopologue $^{13}$CO$_2$ and \citet{Woitke2018} showed how multiple species can be used to determine the C/O ratio in disks \citep[see also][]{Anderson2021}. The models can also be used to generate a testbed for retrieval codes (such as the slab models and CLicK introduced above). Figure~\ref{fig:AATau_simulation} shows the predicted mid-IR spectrum for the DIANA SED model of AA\,Tau\footnote{DIANA models are publically available from https://prodimo.iwf.oeaw.ac.at/models.}. We used here a gas-to-dust ratio of 1000 and a selection of the molecular data listed in Table~\ref{tab:molecular-data}. The disk model is re-calculated with {\sc ProDiMo}, using the large DIANA chemical network \citep{Kamp2017}. We then post-process the results with FLiTs to generate a mid-IR spectrum with a resolution of 30\,000. We add a typical noise level of 300 (representative for what MINDS aims at for T\,Tauri~disks) and sample with a constant resolution of 3000 (oversample$\!=\!2$). The model spectrum matches well the Spitzer IRS high resolution spectrum (Fig.~\ref{fig:AATau_simulation}); the continuum flux offset at $15~\mu$m is only 0.05~Jy. It is clear that we need high S/N MIRI spectra (S/N$\gg$\,200) to be able to reveal for example the weak R- and P-branch lines in the \ce{CO2} spectrum.

\section{First results}
\label{sec:results}

We present here a brief summary of the first results from MINDS, before focussing on one specific object, TW\,Hya, in more detail.

\subsection{Early MINDS results}
\label{sec:early-results}

Within the MINDS program we have presently observed more than 30 T\,Tauri stars with different disk sizes and masses and a variety of disk substructures (see Table~\ref{tab:sample}). This will allow us to constrain statistically for example the dependence of the water distribution as a function of disk properties. In the disk around the T\,Tauri star GW\,Lup, we discovered the molecular emission from CO, H$_2$O, OH, CO$_2$, and HCN \citep{Grant2023}. For the first time, we could also detect the $^{13}$CO$_2$ isotopologue in a disk. We derived a surprisingly high ratio of CO$_2$-to-H$_2$O column density of $\sim\,0.7$. This high ratio may be caused by an inner cavity between the H$_2$O and CO$_2$ mid-plane snowlines, blocking the transport of water-ice rich pebbles \citep{Vlasblom2023}. Another example of the MINDS program is the observation of the T\,Tauri star Sz\,98 with its unusually large dust disk \citep{Gasman2023}. Despite the presence of such a large disk, we find abundant water emission in the inner disk, pointing to a subsolar C/O ratio. This is in strong contrast to the outer disk where a C/O ratio larger than unity was measured with ALMA  \citep{Miotello2019}. A MINDS study by \cite{Perotti2023} found water in the very inner disk of the young planetary system PDS\,70. The water molecules can be both self-shielded and protected by the presence of small micron-sized dust particles. The water reservoir could be formed in situ in the inner disk or be sustained by water-rich dust particles transported from the outer disk. These scenarios are not exclusive and plausibly both contribute. The two planets and evidence for accretion flows have been discovered in the MINDS NIRCAM 
images of PDS\,70 (Christiaens et al.\ submitted).

The MINDS sample contains also a number of disks around very low-mass M-type stars dwarfs. In the disk around such a low-mass star (J16053215) the emission of a variety of hydrocarbons, including benzene was found \citep{Tabone2023}. Strong broad bumps at 7.7 and 13.7~$\mu$m are due to optically thick \ce{C2H2} emission. Its column densities are a facor $\sim\!1000$ higher than in other disks leading to a pseudo-continuum. The JWST observations of this source show a complete lack of silicate emission features, indicating that the grains are settled toward the mid-plane or grown to larger than a few micron. The hydrocarbon chemistry points to a large C/O ratio in the inner 0.1~au of this disk \citep[][Kanwar et al.\ submitted]{Kanwar2024}. A possible hypothesis is that oxygen is locked-up in icy pebbles and planetesimals outside of the water ice-line. Another possibility is the faster transport of pebbles covered by water ice causing most oxygen to have been accreted onto the star \citep{Mah2023}.

\subsection{The nearby disk around TW\,Hya} 
\label{sec:TWHya}

The TW\,Hya star-disk system has been intensively studied because of its proximity, accretion behavior, and disk structure. With a Gaia DR3 distance of only 60~pc \citep{Gaia2021}, TW\,Hya is the nearest T\,Tauri star with a dusty gas-rich disk. It is a member of the TWA association with an age of 10$\,\pm 2$~Myr \citep[Gaia DR3,][]{Luhman2023} in agreement with earlier age estimates \citep{Barrado2006,Hoff1998}. Despite this age, the disk is still relatively massive \citep[$\gtrsim\,0.05$~M$_\odot$,][]{Bergin2013,Franceschi2022}. The system is seen nearly face-on \citep{Qi2004} making it an excellent target for infrared spectroscopy of the disk. \citet{Zhang2013} found the Herschel and Spitzer water line detections to be consistent with the inner few au of this system being depleted in water (factor of $\sim\!100$). In a follow-up study \citet{Bosman2019} combined archival Spitzer and VLT/CRIRES data and performed a detailed analysis with a thermochemical model. They concluded that the inner disk is not enriched by ice-covered inward drifting pebbles because the elemental carbon and oxygen abundances are about a factor of 50 smaller than in the interstellar medium \citep[also in agreement with atomic line analysis by][]{McClure2020}. The authors proposed that the drifting pebbles are stopped in a dust trap outside the water ice line. 

The disk around TW\,Hya is characterized by a system of rings and gaps with a high degree of azimuthal symmetry. Based on the analysis of the spectral energy distribution with its small near-infrared excess, the TW\,Hya disk has been classified as a `transition disk' with an characteristic radius of $\sim\!4$~au separating optically thick submicron-sized grains and optically thin dust at smaller radii \citep{Calvet2002}. In scattered light near-infrared images tracing the distribution of submicron-sized particles, three radial gaps have been discovered \citep{vanBoekel2017}. ALMA observations \citep{Andrews2016, Tsukagoshi2016}, revealing the spatial distribution of millimeter-sized dust grains, demonstrate the presence of a system of concentric ring-shaped substructures at the same position. Relevant for the MIRI observations of this disk, we note that the ALMA observations found evidence for the presence of a narrow dark annulus at 1~au. The rim of the innermost (optically thin) dusty disk is located at $\sim\!0.04$~au from the source \citep{GRAVITY2020, GRAVITY2021, GRAVITY2023}. A comprehensive analysis by \citet{Menu2014}, combining mid-infrared interferometry data with sub-millimeter and eVLA observations provided a structural model of the inner disk regions consisting of a peak surface density at about 2.5~au, with a smooth decrease inwards to 0.35~au. 

\begin{figure}
    \centering
    \includegraphics[width=0.6\textwidth]{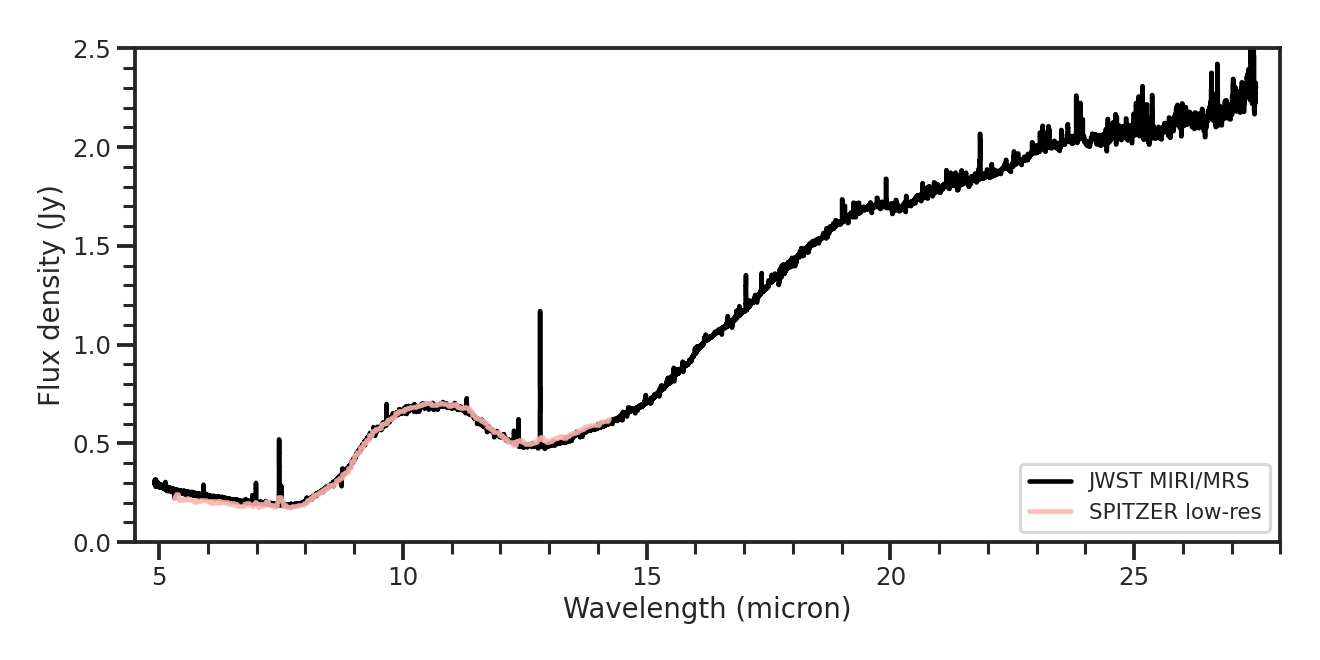}\\[-4mm]
    \caption{Comparison of JWST/MIRI and Spitzer low resolution spectrum of TW\,Hya (no scaling).}
    \label{fig:TWHya-Spitzer-MIRI}
\end{figure}

\begin{figure}
    \centering
   \includegraphics[width=0.87\textwidth]{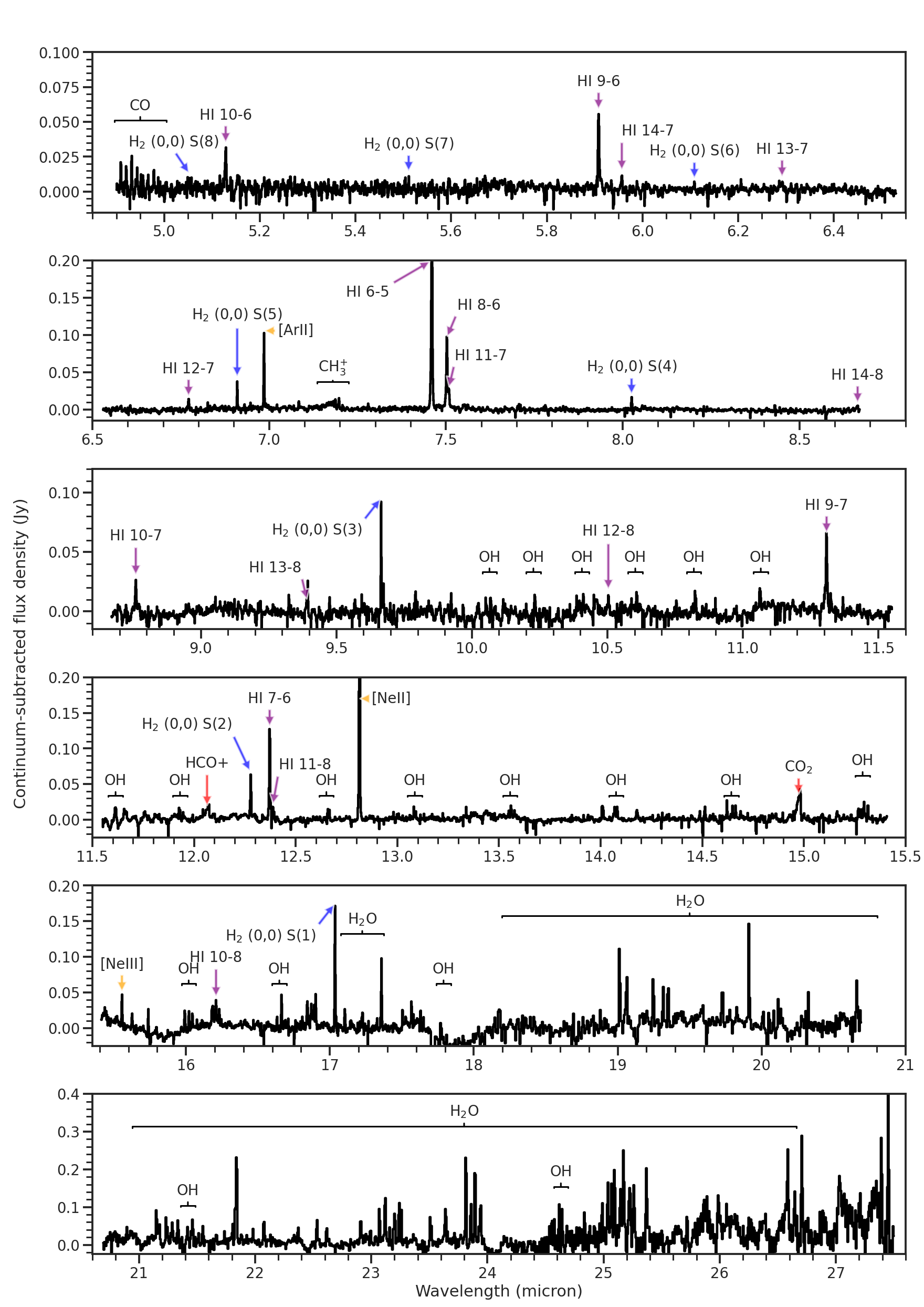}
    \caption{Continuum subtracted TW\,Hya MIRI spectrum with line identifications.}
    \label{fig:TWHya-cont-sub}
\end{figure}

\subsubsection{Analysis of the MIRI spectrum}

In Fig.~\ref{fig:TWHya-Spitzer-MIRI} we show the broad-wavelength MIRI spectrum between 4.9 and 27.9~$\mu$m and compare it with the Spitzer low-resolution spectrum. 
The small remaining absorption features are not real and are expected to vanish with improvements in the data reduction pipeline. 
At variance with other sources of the sample \citep[e.g., GW\,Lup, Sz\,98, SY\,Cha; see][]{Grant2023, Gasman2023, Schwarz2023}, within the calibration uncertainties, we do not find any continuum flux difference between Spitzer and JWST up to $15~\mu$m, nor a change in the silicate emission feature shape. We leave the study of potential line variability to future work.
We define a continuum based on line-free regions and subtract this continuum to identify emission features in the spectrum. Figure~\ref{fig:TWHya-cont-sub} zooms into the spectrum and identifies the key lines and features. 

\underline{Fine-structure lines}: At wavelengths below $10\,\mu$m, we detect the [Ar\,{\sc ii}] emission line at 6.99~$\mu$m. The [Ar\,{\sc iii}] emission lines at 6.36, 8.99 and $21.82~\mu$m are not detected.
We also detect [Ne\,{\sc ii}] at $12.81~\mu$m and [Ne\,{\sc iii}] at $15.56~\mu$m \citep[previously detected by Spitzer,][]{Najita2010}. 

\underline{H$_2$ lines:} The full series of H$_2$ lines from S(8) to S(3) is detected in addition to the previously reported S(2) and S(1) lines seen in Spitzer high resolution spectra \citep{Najita2010}. 

\underline{HD lines:} Several of the HD lines in the MIRI wavelength range are blended with water lines. The longer wavelength lines at $15.022$ and $23.033~\mu$m are not detected above the noise level. We find that the former one coincides with a \ce{CO2} P-branch line and ruling out a detection with high confidence requires an accurate fitting of the \ce{CO2} molecule emission.

\begin{figure}
    \centering
    \includegraphics[width=1.0\textwidth]{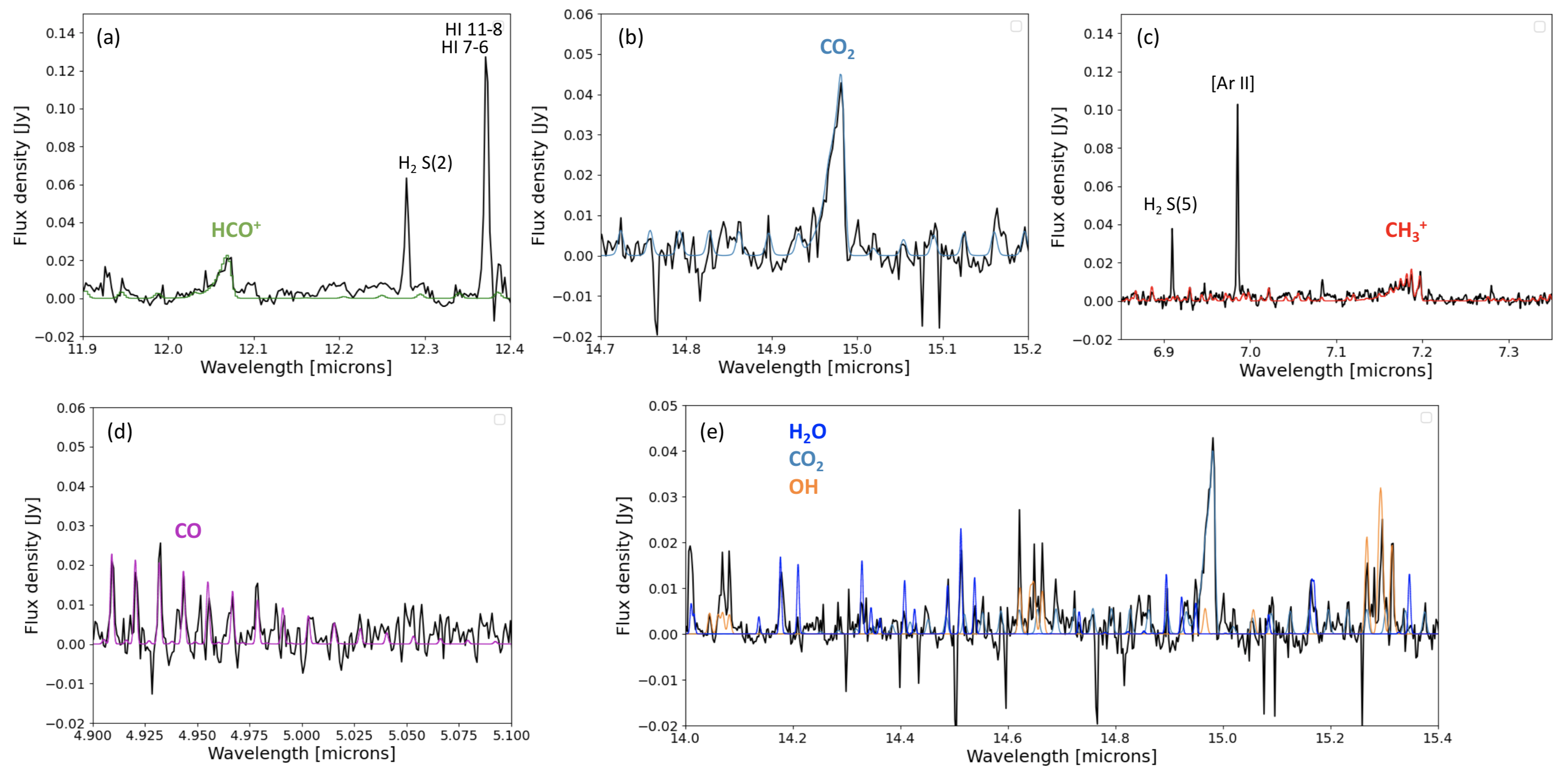}\\[-2mm]
    \caption{Continuum subtracted TW\,Hya MIRI spectrum (black) with (a) \ce{HCO+} slab model (green) overplotted ($T=250$~K, also visible the \ce{H2} S(2) line at 12.28~$\mu$m and the H\,{\sc i} 11-8 and 7-6 lines at 12.39 and 12.37~$\mu$m), (b) \ce{CO2} slab model (blue) overplotted ($T=200$~K), (c) \ce{CH3+} slab model (red) overplotted ($T=500$~K, also visible the \ce{H2} S(5) line at 6.91~$\mu$m and the \ce{Ar+} fine-structure line at 6.99~$\mu$m), (d) CO slab model (magenta) overplotted ($T=500$~K), (e) \ce{H2O}, OH and \ce{CO2} slab models (colors indicated in legend, $T=400, 1000, 200$~K, respectively) in the $14.0-15.4~\mu$m wavelength range. All slab models are calculated in LTE and used only for identification of molecular emission features.}
    \label{fig:TWHya-zoom-molecular-spectra}
\end{figure}

\underline{OH and water:} We detect OH emission longward of $\sim\!9~\mu$m and water emission longward of $14~\mu$m. The shortest water emission so far reported for TW\,Hya has been at $21.85~\mu$m \citep{Bosman2019}; we now clearly detect a number of water lines also in the 14-15~$\mu$m region (Fig.~\ref{fig:TWHya-zoom-molecular-spectra}e), but not the \ce{H2O} ro-vibrational lines at 6-7~$\mu$m. Between 9 and 10.5~$\mu$m, we see the clear signature of OH prompt emission \citep[UV photodissociation of water populates only two of the four OH components,][]{Tabone2021, Zannese2023}. The OH quadruplets beyond $14$~$\mu$m are more symmetric than those at shorter wavelengths (see Fig.~\ref{fig:TWHya-zoom-molecular-spectra}e). This could indicate either (1) collisional quenching, (2) a radiative cascade of rotationally excited OH produced by water photodissociation, or (3) chemical pumping via the reaction O~+~\ce{H2}, which populates rotational levels up to $J\!\sim\!20$ ($\lambda\!\gtrsim\!16~\mu$m). 

\underline{Molecular ions}: The ro-vibrational band of \ce{CH3+} at 7.15~$\mu$m is clearly detected (Fig.~\ref{fig:TWHya-zoom-molecular-spectra}c); this emission band has recently been detected for the first time by the PDRs4All team in the spectrum of a proplyd in Orion \citep{Berne2023} and molecular data has been compiled by \citet{Changala2023}. A tentative detection corresponding to the Q-branch of \ce{HCO+} had been noted by \citet{Najita2010} and rotational line emission is clearly detected with the SMA inside 1\arcsec\ but offset to the center \citep[inside $\sim\,60$~au,][]{Cleeves2015}. The MIRI spectrum now resolves the Q-branch of \ce{HCO+} at 12.1~$\mu$m and tentatively detects in addition several P- and R-branch lines next to it (Fig.~\ref{fig:TWHya-zoom-molecular-spectra}a, molecular slab spectra from F.\ Helmich 1996, PhD thesis).

\underline{Other molecules:} The high-$J$ lines of the P-branch of the CO ro-vibrational band ($v\!=\!1-0$) are well detected from P25 to P31 (Fig.~\ref{fig:TWHya-zoom-molecular-spectra}d); future data reduction improvements such as better fringe correction are warranted to quantitatively model the CO emission and derive upper limits for $^{13}$CO. The Q-branch of CO$_2$ is now resolved at the MIRI spectral resolution and several P- and R-branch lines are also detected (Fig.~\ref{fig:TWHya-zoom-molecular-spectra}b). HCN, \ce{C2H2}, \ce{NH3}, \ce{CH3} and \ce{CH4} are not detected above the current noise level. Given the high S/N of our spectrum, we can now put more stringent upper limits on those molecules and we will come back to this in the next section.

\subsubsection{Confronting the MIRI spectrum with a thermo-chemical disk model}
\label{Sec:Thermo-chemical-Model}

\begin{figure}
    \centering
    \includegraphics[width=1.0\textwidth]{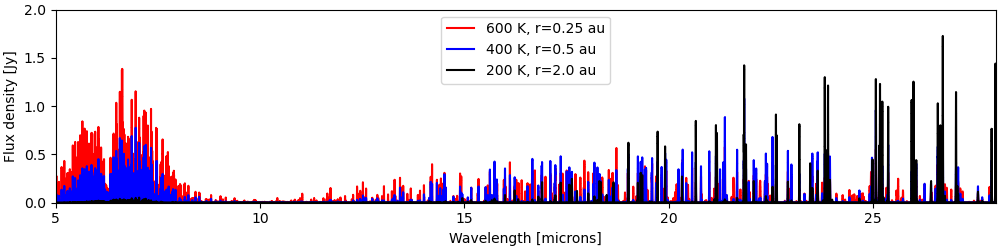}
    \caption{LTE water slab models for a column density of $4\times 10^{18}$~cm$^{-2}$ (using a canonical o/p ratio of 3) and different temperatures and emitting areas $\pi r^2$ (as indicated in legend).}
    \label{fig:TWHyamodel-water-slab}
\end{figure}

So far, we have mostly used 0D slab models to guide the quantitative analysis of the MIRI spectra. These simple models are extremely powerful to build up a first level of understanding. For example, slab models show that if there were trace amounts of warm water ($N({\rm \ce{H2O}})\!=\!4 \times 10^{18}$~cm$^{-2}$, $T\!=\!400-600$~K) in the small inner disk of TW\,Hya ($\lesssim\,1$~au), we would have easily detected that in our JWST/MIRI spectrum, especially at 5-8~$\mu$m (Fig.~\ref{fig:TWHyamodel-water-slab}).

\begin{figure}
    \centering
\includegraphics[width=1.0\textwidth]{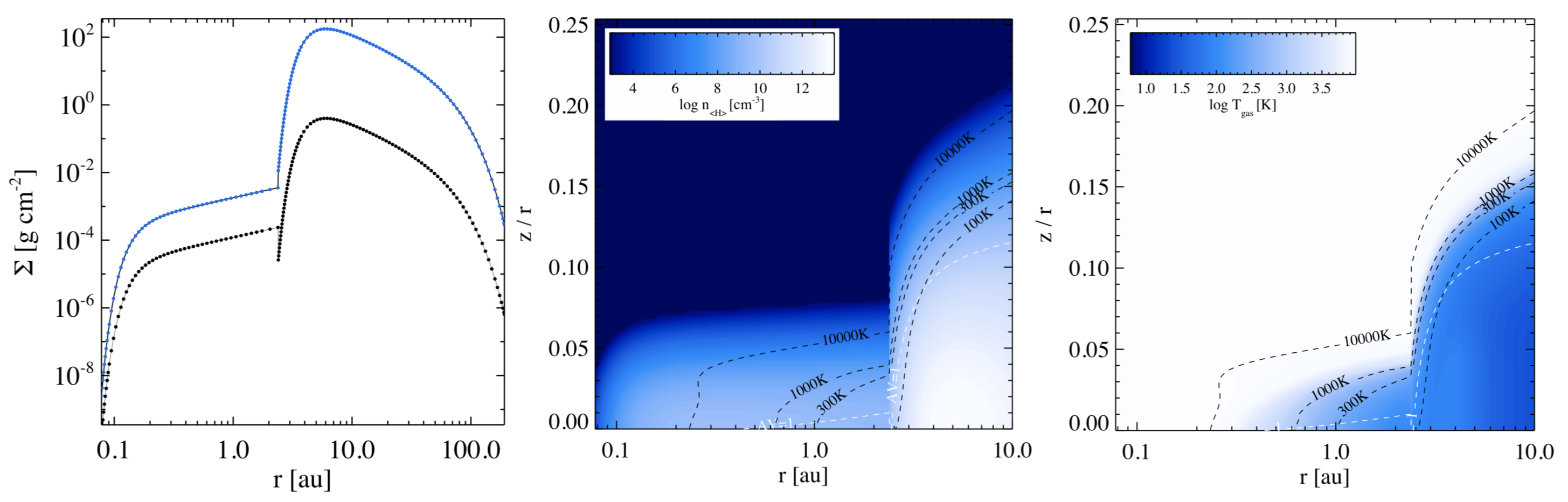}\\[-2mm]
    \caption{The radial gas (blue) and dust (black) surface density profile across the entire disk (left panel) and the density distribution (middle panel), gas temperature (right panel) inside 10~au in the adjusted TW\,Hya DIANA thermo-chemical disk model (elemental abundances of C, N, O depleted by a factor 50 in the inner 2.4~au). Overplotted are the gas temperature contours of 100, 300, 1000 and 10000~K (black) and the $A_{\rm V}\!=\!1$~mag line (white).}
    \label{fig:TWHyamodel-structure}
\end{figure}

In the following, we use the DIANA TW\,Hya disk thermo-chemical model \citep{Woitke2019} to put a few of the key findings in the spectrum into a larger perspective. The DIANA project used the UV/optical/IR/submm photometry as well as Spitzer and Herschel/SPIRE spectra, 57 emission line fluxes, the ALMA $850~\mu$m radial continuum emission profile and the CO $J\!=\!3-2$ line profile to constrain the dust+gas disk structure of this object through an evolutionary fitting strategy. The DIANA disk model has been updated slightly for this work and re-run using ProDiMo. The chemical network is the large DIANA network plus D, D$^+$, HD and HD$^+$ with 239 species and 3147 reactions compiled by \citet{Kamp2017} and updated by \citep{Kanwar2024}. We changed the inner rim of the outer disk to be at 2.4~au \citep{vanBoekel2017} and adjusted the scale height to $0.06$~au at 4~au and the flaring to $\beta\!=\!1.05$ to keep the agreement with the overall observed SED. Figure~\ref{fig:TWHyamodel-structure} shows the new radial gas surface density profile, and the 2D distribution of the gas density\footnote{$n_{\langle {\rm H} \rangle}$ is the total hydrogen number density $n({\rm H})+2 n({\rm H_2})$. The PDR parameter $\chi$ is the UV radiation field integrated between 912 and 2050~\AA\ normalized to that of \citet{Draine1978}}. and temperature for the inner 10~au. This model does not capture the radial substructure detected by SPHERE and ALMA; however, this substructure represents rather a moderate modulation of the surface density and based on previous model refinements, we do not expect this to affect much the outer disk results \citep[e.g.,][]{Muro-Arena2018,Rab2020}.

\begin{figure}
    \centering
    \vspace*{-4mm}
\includegraphics[width=1.0\textwidth]{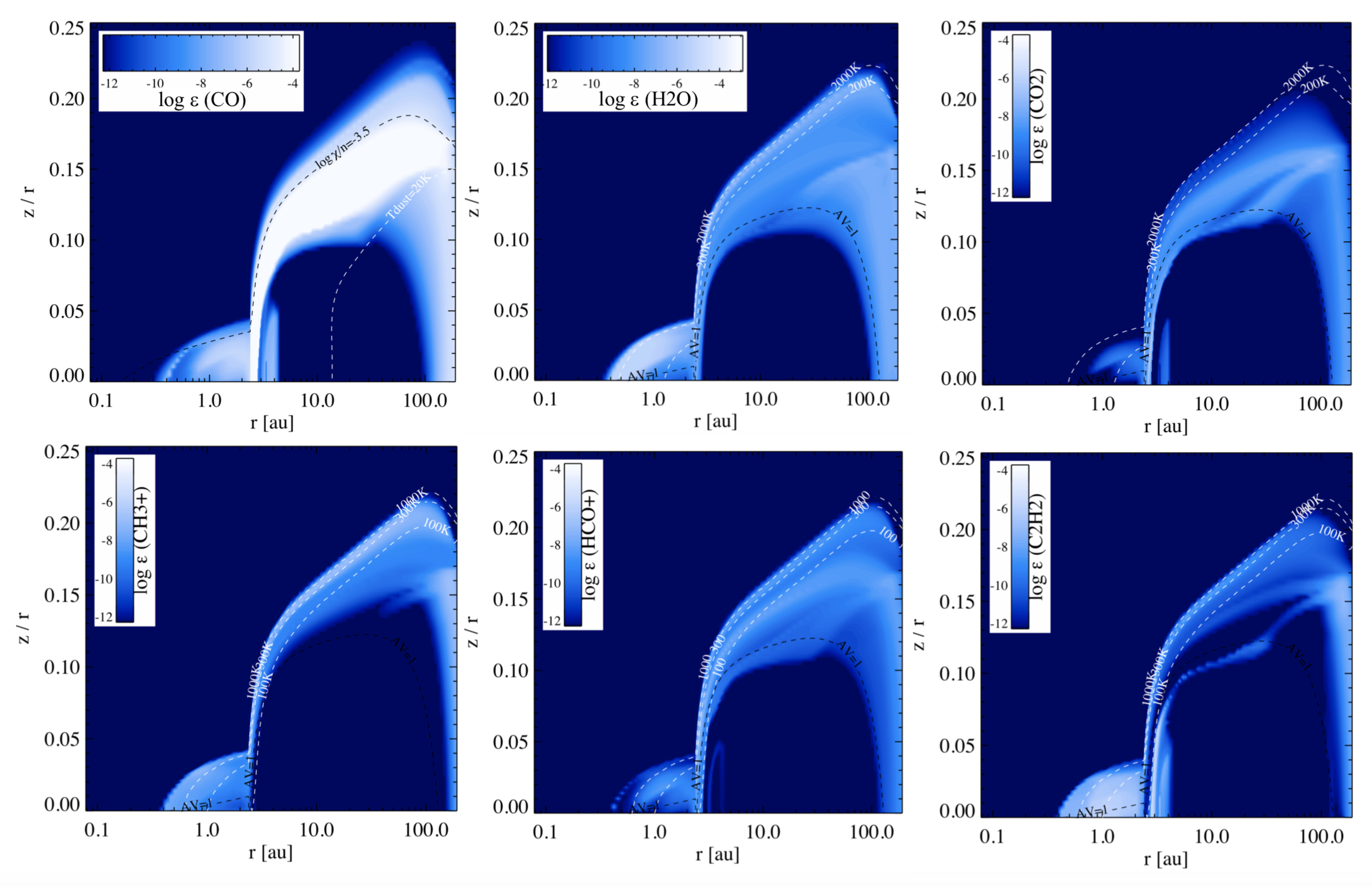}\\[-3mm]
    \caption{Selected abundance distribution of specific molecules and molecular ions in the adjusted TW\,Hya DIANA thermo-chemical disk model (elemental abundances of C, N, O depleted by a factor 50 in the inner 2.4~au). Overplotted are the PDR parameter $\log \chi/n_{\langle \rm H \rangle}$ (see footnote in Sect.~\ref{Sec:Thermo-chemical-Model}) and the dust temperature of 20~K for CO, and for all other molecules the gas temperature contours of 100, 300 and 1000~K (white) and the $A_{\rm V}\!=\!1$~mag line (black).}
    \label{fig:TWHyamodel-abundances}
\end{figure}

We introduced a `rounded' inner rim \citep[gradual increase of column density towards the inner edge of the outer disk,][]{Woitke2023} to have a radially `extended' region where the \ce{HCO+} and \ce{CH3+} molecular ions are abundant (Fig.~\ref{fig:TWHyamodel-abundances}). These molecular ions reside in the transition regime from optically thin to thick (like in PDRs) and those regions would be very geometrically thin if the column density builds up very abruptly over radius. It is interesting to note that such a `rounded' rim was also required to fit the infrared dust interferometry data \citep{Menu2014}.

\begin{figure}
    \hspace*{-10mm}\includegraphics[width=1.1\textwidth]{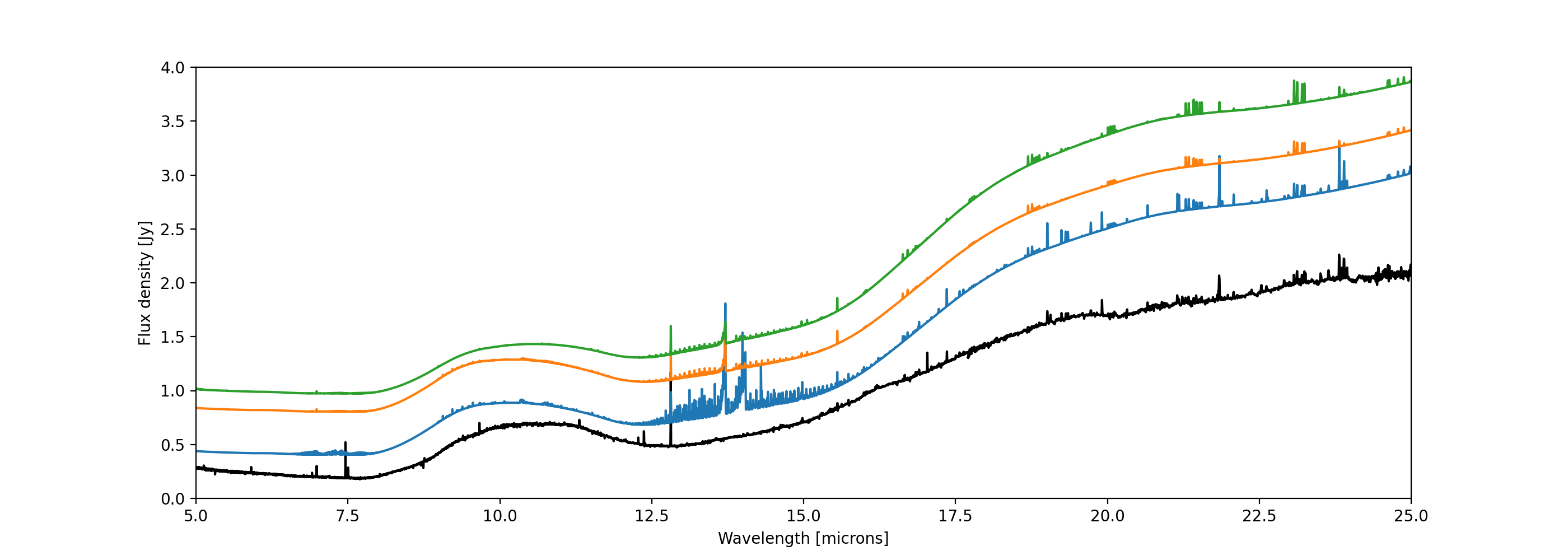}
    \caption{JWST/MIRI MRS spectrum (black) compared to the adjusted DIANA TW\,Hya model using a gas-to-dust mass ratio in the inner disk of 15 (blue, shifted by 0.2\,Jy), and 0.15 (green, shifted by 1\,Jy). The orange spectrum (shifted by 0.6\,Jy) shows the model with a gas-to-dust mass ratio of 15 and the C, N and O abundances in the inner 2.4~au lowered by a factor 50.}
    \label{fig:TWHya-obs-model}
\end{figure}

The \ce{H2}, CO ro-vib, and fine-structure lines of noble gases originate predominantly from the outer disk in our model; the \ce{H2O} lines originate predominantly from inside 2.4~au. \ce{NH3}, HCN, \ce{C2H2} and \ce{CO2} emission in our model partially originate from the inner 2.4~au and partially from the inner rim of the outer disk (Fig.~\ref{fig:TWHyamodel-abundances}). Our standard adjusted DIANA model (low ISM elemental abundances\footnote{These abundances refer to the gas phase abundances derived for the ISM, see Table~5 of \cite{Kamp2017}.}) still overpredicts those features and shows some water emission at $\sim\!7~\mu$m (Fig.~\ref{fig:TWHya-obs-model}, blue spectrum) undetected in our MIRI data. 

Hence, we explore an elemental abundance depletion of C, N and O by a factor 50 for the inner 2.4~au following \citet{Bosman2019} and \citet{McClure2020}. Reducing the elemental abundances of C and O also in the outer disk leads, in our model, to water (at longer wavelengths) and also CO emission line fluxes which are too weak compared to the observations. Such an elemental depletion in the inner disk indeed leads to much weaker \ce{CO2} emission compared to the standard model (originating now in the inner rim of the outer disk), but the \ce{C2H2} is still too strong (Fig.~\ref{fig:TWHya-obs-model}, orange spectrum, and Fig.~\ref{fig:TWHyamodel-abundances}); the bulk of the \ce{C2H2} emission originates inside 2.4~au despite the low C abundance.

An alternative scenario explaining the absence of water, HCN and \ce{C2H2} emission could be an overall lack of gas inside 2.4~au. However, if we lower the gas mass of the inner disk in the thermo-chemical model, turbulence can no longer sustain small grains in the disk surface and the $10~\mu$m silicate feature weakens substantially (and also the continuum due to stronger settling, Fig.~\ref{fig:TWHya-obs-model}, green spectrum). Our modeling approach includes the self-consistent vertical dust settling \citep{Dubrulle1995}, and so the decreasing gas density removes the pressure support for the remaining small silicate grains. 
If we use a gas-to-dust ratio of $0.15$ instead of $15$ (adjusted DIANA model), very little molecular gas remains present inside $\sim\,2.4$~au. However, we still get weak features of \ce{C2H2} originating now from beyond 2.4~au; water, HCN and \ce{CO2} emission disappears. To keep a strong silicate feature in the absence of gas, we could still adjust the grain size distribution, composition and/or scale height. In addition, the spatial distribution of small grains could be de-coupled from the larger grains due to gap filtration effects and/or the presence of planets/planetesimals. 

Given the above experiments of comparing a series of thermo-chemical disk models with varying gas-to-dust mass ratios and elemental abundances in the inner disk to the observed TW\,Hya MIRI spectrum, we prefer the explanation of an inner disk ($r\!<\!2.4$~au) that contains some remnant gas with lower than ISM elemental abundances and which has a 'smooth' transition from the inner to the outer disk. An exploration of a much larger parameter space including also alternative physical scenarios for the dust distribution is left for a future paper.

\section{Outlook} \label{sec:outlook}

The MINDS data presents an incredibly rich set of spectra and images which likely holds many new discoveries yet to come. One immediate example is the search for protoplanets and study of resolved disk emission in TW\,Hya. 

More generally, we are now starting to search for extended line emission across the entire MINDS sample. With time, our knowledge of the instrumental PSF of MIRI/MRS will improve and we can also gather a library of observed PSFs, composed of the unresolved sources of our sample and the archive. With this in hand, the entire MIRI/MRS data could be searched for extended emission using similar strategies as the ones implemented in \citet{Ruffio2023} for NIRSpec or using PSF subtraction via a PCA approach (also common in high-contrast imaging). 

Besides the individual object studies outlined in Sect.~\ref{sec:early-results}, we will study the broader science questions outlined in Sect.~\ref{sec:goals} on well-defined sub-samples of the MINDS sources and also use the synergy with ALMA data, interferometric data, near-IR imaging and ground-based high-spectral resolution spectra. In addition, our modeling will gradually move from simple LTE 0D slab models to more complex retrieval tools (e.g.\ CLicK) and grids of forward disk models to support the interpretation of our data.

\section{Acknowledgements}

This work is based on observations made with the NASA/ESA/CSA James Webb Space Telescope. The data were obtained from the Mikulski Archive for Space Telescopes at the Space Telescope Science Institute, which is operated by the Association of Universities for Research in Astronomy, Inc., under NASA contract NAS 5-03127 for JWST. These observations are associated with the European MIRI GTO program MINDS, program \#1282. I.K.\ acknowledges financial support and the hospitality of the Max Planck Institute for Astronomy in Heidelberg during her visit February-June 2023, where most of this article was compiled.

The following National and International Funding Agencies funded and supported the MIRI development: NASA; ESA; Belgian Science Policy Office (BELSPO); Centre Nationale d’Etudes Spatiales (CNES); Danish National Space Centre; Deutsches Zentrum f\"{u}r Luft- und Raumfahrt (DLR); Enterprise Ireland; Ministerio de Econom\'ia y Competividad; Netherlands Research School for Astronomy (NOVA); Netherlands Organisation for Scientific Research (NWO); Science and Technology Facilities Council; Swiss Space Office; Swedish National Space Agency; and UK Space Agency.
A.C.G. acknowledges from PRIN-MUR 2022 20228JPA3A “The path to star and planet formation in the JWST era (PATH)” funded by NextGeneration EU and by INAF-GoG 2022 “NIR-dark Accretion Outbursts in Massive Young stellar objects (NAOMY)” and Large Grant INAF 2022 “YSOs Outflows, Disks and Accretion: towards a global framework for the evolution of planet forming systems (YODA).
E.v.D. acknowledges support from the ERC grant 101019751 MOLDISK and the Danish National Research Foundation through the Center of Excellence ``InterCat'' (DNRF150). 
T.H. and K.S. acknowledge support from the European Research Council under the Horizon 2020 Framework Program via the ERC Advanced Grant Origins 83 24 28. 
I.K., A.M.A., and E.v.D. acknowledge support from grant TOP-1 614.001.751 from the Dutch Research Council (NWO).
I.K. and J.K. acknowledge funding from H2020-MSCA-ITN-2019, grant no. 860470 (CHAMELEON).
B.T. is a Laureate of the Paris Region fellowship program, which is supported by the Ile-de-France Region and has received funding under the Horizon 2020 innovation framework program and Marie Sklodowska-Curie grant agreement No. 945298.
O.A. and V.C. acknowledge funding from the Belgian F.R.S.-FNRS.
I.A., D.G. and B.V. thank the Belgian Federal Science Policy Office (BELSPO) for the provision of financial support in the framework of the PRODEX Programme of the European Space Agency (ESA).
L.C. acknowledges support by grant PIB2021-127718NB-I00,  from the Spanish Ministry of Science and Innovation/State Agency of Research MCIN/AEI/10.13039/501100011033.
T.P.R acknowledges support from ERC grant 743029 EASY.
D.R.L. acknowledges support from Science Foundation Ireland (grant number 21/PATH-S/9339).
D.B. and M.M.C. have been funded by Spanish MCIN/AEI/10.13039/501100011033 grants PID2019-107061GB-C61 and No. MDM-2017-0737. 
M.T. acknowledges support from the ERC grant 101019751 MOLDISK.

\bibliography{references}{}

\begin{thebibliography}{}
\expandafter\ifx\csname natexlab\endcsname\relax\def\natexlab#1{#1}\fi
\providecommand{\url}[1]{\href{#1}{#1}}
\providecommand{\dodoi}[1]{doi:~\href{http://doi.org/#1}{\nolinkurl{#1}}}
\providecommand{\doeprint}[1]{\href{http://ascl.net/#1}{\nolinkurl{http://ascl.net/#1}}}
\providecommand{\doarXiv}[1]{\href{https://arxiv.org/abs/#1}{\nolinkurl{https://arxiv.org/abs/#1}}}

\bibitem[{{ALMA Partnership} {et~al.}(2015){ALMA Partnership}, {Brogan},
  {P{\'e}rez}, {Hunter}, {Dent}, {Hales}, {Hills}, {Corder}, {Fomalont},
  {Vlahakis}, {Asaki}, {Barkats}, {Hirota}, {Hodge}, {Impellizzeri}, {Kneissl},
  {Liuzzo}, {Lucas}, {Marcelino}, {Matsushita}, {Nakanishi}, {Phillips},
  {Richards}, {Toledo}, {Aladro}, {Broguiere}, {Cortes}, {Cortes}, {Espada},
  {Galarza}, {Garcia-Appadoo}, {Guzman-Ramirez}, {Humphreys}, {Jung}, {Kameno},
  {Laing}, {Leon}, {Marconi}, {Mignano}, {Nikolic}, {Nyman}, {Radiszcz},
  {Remijan}, {Rod{\'o}n}, {Sawada}, {Takahashi}, {Tilanus}, {Vila Vilaro},
  {Watson}, {Wiklind}, {Akiyama}, {Chapillon}, {de Gregorio-Monsalvo}, {Di
  Francesco}, {Gueth}, {Kawamura}, {Lee}, {Nguyen Luong}, {Mangum}, {Pietu},
  {Sanhueza}, {Saigo}, {Takakuwa}, {Ubach}, {van Kempen}, {Wootten},
  {Castro-Carrizo}, {Francke}, {Gallardo}, {Garcia}, {Gonzalez}, {Hill},
  {Kaminski}, {Kurono}, {Liu}, {Lopez}, {Morales}, {Plarre}, {Schieven},
  {Testi}, {Videla}, {Villard}, {Andreani}, {Hibbard}, \&
  {Tatematsu}}]{ALMA2015}
{ALMA Partnership}, {Brogan}, C.~L., {P{\'e}rez}, L.~M., {et~al.} 2015, \apjl,
  808, L3, \dodoi{10.1088/2041-8205/808/1/L3}

\bibitem[{{Anderson} {et~al.}(2021){Anderson}, {Blake}, {Cleeves}, {Bergin},
  {Zhang}, {Schwarz}, {Salyk}, \& {Bosman}}]{Anderson2021}
{Anderson}, D.~E., {Blake}, G.~A., {Cleeves}, L.~I., {et~al.} 2021, \apj, 909,
  55, \dodoi{10.3847/1538-4357/abd9c1}

\bibitem[{{Andrews}(2020)}]{Andrews2020}
{Andrews}, S.~M. 2020, \araa, 58, 483,
  \dodoi{10.1146/annurev-astro-031220-010302}

\bibitem[{{Andrews} {et~al.}(2016){Andrews}, {Wilner}, {Zhu}, {Birnstiel},
  {Carpenter}, {P{\'e}rez}, {Bai}, {{\"O}berg}, {Hughes}, {Isella}, \&
  {Ricci}}]{Andrews2016}
{Andrews}, S.~M., {Wilner}, D.~J., {Zhu}, Z., {et~al.} 2016, \apjl, 820, L40,
  \dodoi{10.3847/2041-8205/820/2/L40}

\bibitem[{{Andrews} {et~al.}(2018){Andrews}, {Huang}, {P{\'e}rez}, {Isella},
  {Dullemond}, {Kurtovic}, {Guzm{\'a}n}, {Carpenter}, {Wilner}, {Zhang}, {Zhu},
  {Birnstiel}, {Bai}, {Benisty}, {Hughes}, {{\"O}berg}, \&
  {Ricci}}]{Andrews2018}
{Andrews}, S.~M., {Huang}, J., {P{\'e}rez}, L.~M., {et~al.} 2018, \apjl, 869,
  L41, \dodoi{10.3847/2041-8213/aaf741}

\bibitem[{{Antonellini} {et~al.}(2015){Antonellini}, {Kamp},
  {Riviere-Marichalar}, {Meijerink}, {Woitke}, {Thi}, {Spaans}, {Aresu}, \&
  {Lee}}]{Antonellini2015}
{Antonellini}, S., {Kamp}, I., {Riviere-Marichalar}, P., {et~al.} 2015, \aap,
  582, A105, \dodoi{10.1051/0004-6361/201525724}

\bibitem[{{Aoyama} {et~al.}(2018){Aoyama}, {Ikoma}, \& {Tanigawa}}]{Aoyama2018}
{Aoyama}, Y., {Ikoma}, M., \& {Tanigawa}, T. 2018, \apj, 866, 84,
  \dodoi{10.3847/1538-4357/aadc11}

\bibitem[{{Argyriou} {et~al.}(2020){Argyriou}, {Wells}, {Glasse}, {Lee},
  {Royer}, {Vandenbussche}, {Malumuth}, {Glauser}, {Kavanagh}, {Labiano},
  {Lahuis}, {Mueller}, \& {Patapis}}]{Argyriou2020}
{Argyriou}, I., {Wells}, M., {Glasse}, A., {et~al.} 2020, \aap, 641, A150,
  \dodoi{10.1051/0004-6361/202037535}

\bibitem[{{Argyriou} {et~al.}(2023){Argyriou}, {Glasse}, {Law}, {Labiano},
  {{\'A}lvarez-M{\'a}rquez}, {Patapis}, {Kavanagh}, {Gasman}, {Mueller},
  {Larson}, {Vandenbussche}, {Glauser}, {Royer}, {Dicken}, {Harkett},
  {Sargent}, {Engesser}, {Jones}, {Kendrew}, {Noriega-Crespo}, {Brandl},
  {Rieke}, {Wright}, {Lee}, \& {Wells}}]{Argyriou2023}
{Argyriou}, I., {Glasse}, A., {Law}, D.~R., {et~al.} 2023, \aap, 675, A111,
  \dodoi{10.1051/0004-6361/202346489}

\bibitem[{{Bae} {et~al.}(2022){Bae}, {Isella}, {Zhu}, {Martin}, {Okuzumi}, \&
  {Suriano}}]{BaePPVII2022}
{Bae}, J., {Isella}, A., {Zhu}, Z., {et~al.} 2022, arXiv e-prints,
  arXiv:2210.13314, \dodoi{10.48550/arXiv.2210.13314}

\bibitem[{{Baldovin-Saavedra} {et~al.}(2012){Baldovin-Saavedra}, {Audard},
  {Carmona}, {G{\"u}del}, {Briggs}, {Rebull}, {Skinner}, \&
  {Ercolano}}]{BaldovinSaavedra2012}
{Baldovin-Saavedra}, C., {Audard}, M., {Carmona}, A., {et~al.} 2012, \aap, 543,
  A30, \dodoi{10.1051/0004-6361/201118329}

\bibitem[{{Barrado Y Navascu{\'e}s}(2006)}]{Barrado2006}
{Barrado Y Navascu{\'e}s}, D. 2006, \aap, 459, 511,
  \dodoi{10.1051/0004-6361:20065717}

\bibitem[{{Benisty} {et~al.}(2015){Benisty}, {Juhasz}, {Boccaletti},
  {Avenhaus}, {Milli}, {Thalmann}, {Dominik}, {Pinilla}, {Buenzli}, {Pohl},
  {Beuzit}, {Birnstiel}, {de Boer}, {Bonnefoy}, {Chauvin}, {Christiaens},
  {Garufi}, {Grady}, {Henning}, {Huelamo}, {Isella}, {Langlois}, {M{\'e}nard},
  {Mouillet}, {Olofsson}, {Pantin}, {Pinte}, \& {Pueyo}}]{Benisty2015}
{Benisty}, M., {Juhasz}, A., {Boccaletti}, A., {et~al.} 2015, \aap, 578, L6,
  \dodoi{10.1051/0004-6361/201526011}

\bibitem[{{Benisty} {et~al.}(2021){Benisty}, {Bae}, {Facchini}, {Keppler},
  {Teague}, {Isella}, {Kurtovic}, {P{\'e}rez}, {Sierra}, {Andrews},
  {Carpenter}, {Czekala}, {Dominik}, {Henning}, {Menard}, {Pinilla}, \&
  {Zurlo}}]{Benisty2021}
{Benisty}, M., {Bae}, J., {Facchini}, S., {et~al.} 2021, \apjl, 916, L2,
  \dodoi{10.3847/2041-8213/ac0f83}

\bibitem[{{Benisty} {et~al.}(2022){Benisty}, {Dominik}, {Follette}, {Garufi},
  {Ginski}, {Hashimoto}, {Keppler}, {Kley}, \& {Monnier}}]{BenistyPPVII2022}
{Benisty}, M., {Dominik}, C., {Follette}, K., {et~al.} 2022, arXiv e-prints,
  arXiv:2203.09991, \dodoi{10.48550/arXiv.2203.09991}

\bibitem[{{Bergin} {et~al.}(2013){Bergin}, {Cleeves}, {Gorti}, {Zhang},
  {Blake}, {Green}, {Andrews}, {Evans}, {Henning}, {{\"O}berg}, {Pontoppidan},
  {Qi}, {Salyk}, \& {van Dishoeck}}]{Bergin2013}
{Bergin}, E.~A., {Cleeves}, L.~I., {Gorti}, U., {et~al.} 2013, \nat, 493, 644,
  \dodoi{10.1038/nature11805}

\bibitem[{{Bern{\'e}} {et~al.}(2023){Bern{\'e}}, {Martin-Drumel}, {Schroetter},
  {Goicoechea}, {Jacovella}, {Gans}, {Dartois}, {Coudert}, {Bergin}, {Alarcon},
  {Cami}, {Roueff}, {Black}, {Asvany}, {Habart}, {Peeters}, {Canin}, {Trahin},
  {Joblin}, {Schlemmer}, {Thorwirth}, {Cernicharo}, {Gerin}, {Tielens},
  {Zannese}, {Abergel}, {Bernard-Salas}, {Boersma}, {Bron}, {Chown},
  {Cuadrado}, {Dicken}, {Elyajouri}, {Fuente}, {Gordon}, {Issa}, {Kannavou},
  {Khan}, {Lacinbala}, {Languignon}, {Le Gal}, {Maragkoudakis}, {Meshaka},
  {Okada}, {Onaka}, {Pasquini}, {Pound}, {Robberto}, {R{\"o}llig}, {Schefter},
  {Schirmer}, {Sidhu}, {Tabone}, {Van De Putte}, {Vicente}, \&
  {Wolfire}}]{Berne2023}
{Bern{\'e}}, O., {Martin-Drumel}, M.-A., {Schroetter}, I., {et~al.} 2023, \nat,
  621, 56, \dodoi{10.1038/s41586-023-06307-x}

\bibitem[{{Bethell} \& {Bergin}(2009)}]{Bethell2009}
{Bethell}, T., \& {Bergin}, E. 2009, Science, 326, 1675,
  \dodoi{10.1126/science.1176879}

\bibitem[{{Birnstiel}(2023)}]{Birnstiel2023}
{Birnstiel}, T. 2023, arXiv e-prints, arXiv:2312.13287.
\newblock \doarXiv{2312.13287}

\bibitem[{{Boccaletti} {et~al.}(2022){Boccaletti}, {Cossou}, {Baudoz},
  {Lagage}, {Dicken}, {Glasse}, {Hines}, {Aguilar}, {Detre}, {Nickson},
  {Noriega-Crespo}, {G{\'a}sp{\'a}r}, {Labiano}, {Stark}, {Rouan}, {Reess},
  {Wright}, {Rieke}, {Garcia Marin}, {Lajoie}, {Girard}, {Perrin}, {Soummer},
  \& {Pueyo}}]{Boccaletti2022}
{Boccaletti}, A., {Cossou}, C., {Baudoz}, P., {et~al.} 2022, \aap, 667, A165,
  \dodoi{10.1051/0004-6361/202244578}

\bibitem[{{Bosman} \& {Banzatti}(2019)}]{Bosman2019}
{Bosman}, A.~D., \& {Banzatti}, A. 2019, \aap, 632, L10,
  \dodoi{10.1051/0004-6361/201936638}

\bibitem[{{Bosman} {et~al.}(2022{\natexlab{a}}){Bosman}, {Bergin}, {Calahan},
  \& {Duval}}]{Bosman2022a}
{Bosman}, A.~D., {Bergin}, E.~A., {Calahan}, J., \& {Duval}, S.~E.
  2022{\natexlab{a}}, \apjl, 930, L26, \dodoi{10.3847/2041-8213/ac66ce}

\bibitem[{{Bosman} {et~al.}(2022{\natexlab{b}}){Bosman}, {Bergin}, {Calahan},
  \& {Duval}}]{Bosman2022b}
{Bosman}, A.~D., {Bergin}, E.~A., {Calahan}, J.~K., \& {Duval}, S.~E.
  2022{\natexlab{b}}, \apjl, 933, L40, \dodoi{10.3847/2041-8213/ac7d9f}

\bibitem[{{Bosman} {et~al.}(2017){Bosman}, {Bruderer}, \& {van
  Dishoeck}}]{Bosman2017}
{Bosman}, A.~D., {Bruderer}, S., \& {van Dishoeck}, E.~F. 2017, \aap, 601, A36,
  \dodoi{10.1051/0004-6361/201629946}

\bibitem[{{Bosman} {et~al.}(2018){Bosman}, {Tielens}, \& {van
  Dishoeck}}]{Bosman2018}
{Bosman}, A.~D., {Tielens}, A. G.~G.~M., \& {van Dishoeck}, E.~F. 2018, \aap,
  611, A80, \dodoi{10.1051/0004-6361/201732056}

\bibitem[{{Brown} {et~al.}(2013){Brown}, {Pontoppidan}, {van Dishoeck},
  {Herczeg}, {Blake}, \& {Smette}}]{Brown2013}
{Brown}, J.~M., {Pontoppidan}, K.~M., {van Dishoeck}, E.~F., {et~al.} 2013,
  \apj, 770, 94, \dodoi{10.1088/0004-637X/770/2/94}

\bibitem[{{Bruderer} {et~al.}(2015){Bruderer}, {Harsono}, \& {van
  Dishoeck}}]{Bruderer2015}
{Bruderer}, S., {Harsono}, D., \& {van Dishoeck}, E.~F. 2015, \aap, 575, A94,
  \dodoi{10.1051/0004-6361/201425009}

\bibitem[{{Bruderer} {et~al.}(2012){Bruderer}, {van Dishoeck}, {Doty}, \&
  {Herczeg}}]{Bruderer2012}
{Bruderer}, S., {van Dishoeck}, E.~F., {Doty}, S.~D., \& {Herczeg}, G.~J. 2012,
  \aap, 541, A91, \dodoi{10.1051/0004-6361/201118218}

\bibitem[{{Burriesci}(2005)}]{Burriesci2005}
{Burriesci}, L.~G. 2005, in Society of Photo-Optical Instrumentation Engineers
  (SPIE) Conference Series, Vol. 5904, Cryogenic Optical Systems and
  Instruments XI, ed. J.~B. {Heaney} \& L.~G. {Burriesci}, 21--29,
  \dodoi{10.1117/12.613596}

\bibitem[{{Bushouse} {et~al.}(2022){Bushouse}, {Eisenhamer}, {Dencheva},
  {Davies}, {Greenfield}, {Morrison}, {Hodge}, {Simon}, {Grumm}, {Droettboom},
  {Slavich}, {Sosey}, {Pauly}, {Miller}, {Jedrzejewski}, {Hack}, {Davis},
  {Crawford}, {Law}, {Gordon}, {Regan}, {Cara}, {MacDonald}, {Bradley},
  {Shanahan}, \& {Jamieson}}]{Bushouse2022}
{Bushouse}, H., {Eisenhamer}, J., {Dencheva}, N., {et~al.} 2022,
  {spacetelescope/jwst: JWST 1.6.2}, 1.6.2, Zenodo,  Zenodo,
  \dodoi{10.5281/zenodo.6984366}

\bibitem[{{Calvet} {et~al.}(2002){Calvet}, {D'Alessio}, {Hartmann}, {Wilner},
  {Walsh}, \& {Sitko}}]{Calvet2002}
{Calvet}, N., {D'Alessio}, P., {Hartmann}, L., {et~al.} 2002, \apj, 568, 1008,
  \dodoi{10.1086/339061}

\bibitem[{{Carr} \& {Najita}(2008)}]{Carr2008}
{Carr}, J.~S., \& {Najita}, J.~R. 2008, Science, 319, 1504,
  \dodoi{10.1126/science.1153807}

\bibitem[{{Cazzoletti} {et~al.}(2018){Cazzoletti}, {van Dishoeck}, {Pinilla},
  {Tazzari}, {Facchini}, {van der Marel}, {Benisty}, {Garufi}, \&
  {P{\'e}rez}}]{Cazzoletti2018}
{Cazzoletti}, P., {van Dishoeck}, E.~F., {Pinilla}, P., {et~al.} 2018, \aap,
  619, A161, \dodoi{10.1051/0004-6361/201834006}

\bibitem[{{Changala} {et~al.}(2023){Changala}, {Chen}, {Le}, {Gans},
  {Steenbakkers}, {Salomon}, {Bonah}, {Schroetter}, {Canin}, {Martin-Drumel},
  {Jacovella}, {Dartois}, {Boy{\'e}-P{\'e}ronne}, {Alcaraz}, {Asvany},
  {Br{\"u}nken}, {Thorwirth}, {Schlemmer}, {Goicoechea}, {Rouill{\'e}},
  {Sidhu}, {Chown}, {Van De Putte}, {Trahin}, {Alarc{\'o}n}, {Bern{\'e}},
  {Habart}, \& {Peeters}}]{Changala2023}
{Changala}, P.~B., {Chen}, N.~L., {Le}, H.~L., {et~al.} 2023, \aap, 680, A19,
  \dodoi{10.1051/0004-6361/202347765}

\bibitem[{{Chen} \& {Szul{\'a}gyi}(2022)}]{Chen2022}
{Chen}, X., \& {Szul{\'a}gyi}, J. 2022, \mnras, 516, 506,
  \dodoi{10.1093/mnras/stac1976}

\bibitem[{{Christiaens} {et~al.}(2023){Christiaens}, {Gonzalez}, {Farkas},
  {Dahlqvist}, {Nasedkin}, {Milli}, {Absil}, {Ngo}, {Cantero}, {Rainot},
  {Hammond}, {Bonse}, {Cantalloube}, {Vigan}, {Kompella}, \&
  {Hancock}}]{Christiaens2023}
{Christiaens}, V., {Gonzalez}, C., {Farkas}, R., {et~al.} 2023, The Journal of
  Open Source Software, 8, 4774, \dodoi{10.21105/joss.04774}

\bibitem[{{Cleeves} {et~al.}(2015){Cleeves}, {Bergin}, {Qi}, {Adams}, \&
  {{\"O}berg}}]{Cleeves2015}
{Cleeves}, L.~I., {Bergin}, E.~A., {Qi}, C., {Adams}, F.~C., \& {{\"O}berg},
  K.~I. 2015, \apj, 799, 204, \dodoi{10.1088/0004-637X/799/2/204}

\bibitem[{{Dent} {et~al.}(2014){Dent}, {Wyatt}, {Roberge}, {Augereau},
  {Casassus}, {Corder}, {Greaves}, {de Gregorio-Monsalvo}, {Hales}, \&
  {Jackson}}]{Dent2014}
{Dent}, W.~R.~F., {Wyatt}, M.~C., {Roberge}, A., {et~al.} 2014, Science, 343,
  1490, \dodoi{10.1126/science.1248726}

\bibitem[{{Draine}(1978)}]{Draine1978}
{Draine}, B.~T. 1978, \apjs, 36, 595, \dodoi{10.1086/190513}

\bibitem[{{Dubrulle} {et~al.}(1995){Dubrulle}, {Morfill}, \&
  {Sterzik}}]{Dubrulle1995}
{Dubrulle}, B., {Morfill}, G., \& {Sterzik}, M. 1995, \icarus, 114, 237,
  \dodoi{10.1006/icar.1995.1058}

\bibitem[{{Dullemond} {et~al.}(2012){Dullemond}, {Juhasz}, {Pohl}, {Sereshti},
  {Shetty}, {Peters}, {Commercon}, \& {Flock}}]{Dullemond2012}
{Dullemond}, C.~P., {Juhasz}, A., {Pohl}, A., {et~al.} 2012, {RADMC-3D: A
  multi-purpose radiative transfer tool}, Astrophysics Source Code Library,
  record ascl:1202.015.
\newblock \doeprint{1202.015}

\bibitem[{{Franceschi} {et~al.}(2022){Franceschi}, {Birnstiel}, {Henning},
  {Pinilla}, {Semenov}, \& {Zormpas}}]{Franceschi2022}
{Franceschi}, R., {Birnstiel}, T., {Henning}, T., {et~al.} 2022, \aap, 657,
  A74, \dodoi{10.1051/0004-6361/202141705}

\bibitem[{{Gaia Collaboration} {et~al.}(2021){Gaia Collaboration}, {Smart},
  {Sarro}, {Rybizki}, {Reyl{\'e}}, {Robin}, {Hambly}, {Abbas}, {Barstow}, {de
  Bruijne}, {Bucciarelli}, {Carrasco}, {Cooper}, {Hodgkin}, {Masana},
  {Michalik}, {Sahlmann}, {Sozzetti}, {Brown}, {Vallenari}, {Prusti},
  {Babusiaux}, {Biermann}, {Creevey}, {Evans}, {Eyer}, {Hutton}, {Jansen},
  {Jordi}, {Klioner}, {Lammers}, {Lindegren}, {Luri}, {Mignard}, {Panem},
  {Pourbaix}, {Randich}, {Sartoretti}, {Soubiran}, {Walton}, {Arenou},
  {Bailer-Jones}, {Bastian}, {Cropper}, {Drimmel}, {Katz}, {Lattanzi}, {van
  Leeuwen}, {Bakker}, {Casta{\~n}eda}, {De Angeli}, {Ducourant}, {Fabricius},
  {Fouesneau}, {Fr{\'e}mat}, {Guerra}, {Guerrier}, {Guiraud}, {Jean-Antoine
  Piccolo}, {Messineo}, {Mowlavi}, {Nicolas}, {Nienartowicz}, {Pailler},
  {Panuzzo}, {Riclet}, {Roux}, {Seabroke}, {Sordo}, {Tanga}, {Th{\'e}venin},
  {Gracia-Abril}, {Portell}, {Teyssier}, {Altmann}, {Andrae}, {Bellas-Velidis},
  {Benson}, {Berthier}, {Blomme}, {Brugaletta}, {Burgess}, {Busso}, {Carry},
  {Cellino}, {Cheek}, {Clementini}, {Damerdji}, {Davidson}, {Delchambre},
  {Dell'Oro}, {Fern{\'a}ndez-Hern{\'a}ndez}, {Galluccio}, {Garc{\'\i}a-Lario},
  {Garcia-Reinaldos}, {Gonz{\'a}lez-N{\'u}{\~n}ez}, {Gosset}, {Haigron},
  {Halbwachs}, {Harrison}, {Hatzidimitriou}, {Heiter}, {Hern{\'a}ndez},
  {Hestroffer}, {Holl}, {Jan{\ss}en}, {Jevardat de Fombelle}, {Jordan},
  {Krone-Martins}, {Lanzafame}, {L{\"o}ffler}, {Lorca}, {Manteiga}, {Marchal},
  {Marrese}, {Moitinho}, {Mora}, {Muinonen}, {Osborne}, {Pancino}, {Pauwels},
  {Recio-Blanco}, {Richards}, {Riello}, {Rimoldini}, {Roegiers}, {Siopis},
  {Smith}, {Ulla}, {Utrilla}, {van Leeuwen}, {van Reeven}, {Abreu Aramburu},
  {Accart}, {Aerts}, {Aguado}, {Ajaj}, {Altavilla}, {{\'A}lvarez}, {{\'A}lvarez
  Cid-Fuentes}, {Alves}, {Anderson}, {Anglada Varela}, {Antoja}, {Audard},
  {Baines}, {Baker}, {Balaguer-N{\'u}{\~n}ez}, {Balbinot}, {Balog}, {Barache},
  {Barbato}, {Barros}, {Bartolom{\'e}}, {Bassilana}, {Bauchet},
  {Baudesson-Stella}, {Becciani}, {Bellazzini}, {Bernet}, {Bertone}, {Bianchi},
  {Blanco-Cuaresma}, {Boch}, {Bombrun}, {Bossini}, {Bouquillon}, {Bragaglia},
  {Bramante}, {Breedt}, {Bressan}, {Brouillet}, {Burlacu}, {Busonero},
  {Butkevich}, {Buzzi}, {Caffau}, {Cancelliere}, {C{\'a}novas},
  {Cantat-Gaudin}, {Carballo}, {Carlucci}, {Carnerero}, {Casamiquela},
  {Castellani}, {Castro-Ginard}, {Castro Sampol}, {Chaoul}, {Charlot},
  {Chemin}, {Chiavassa}, {Cioni}, {Comoretto}, {Cornez}, {Cowell}, {Crifo},
  {Crosta}, {Crowley}, {Dafonte}, {Dapergolas}, {David}, {David}, {de Laverny},
  {De Luise}, {De March}, {De Ridder}, {de Souza}, {de Teodoro}, {de Torres},
  {del Peloso}, {del Pozo}, {Delgado}, {Delgado}, {Delisle}, {Di Matteo},
  {Diakite}, {Diener}, {Distefano}, {Dolding}, {Eappachen}, {Edvardsson},
  {Enke}, {Esquej}, {Fabre}, {Fabrizio}, {Faigler}, {Fedorets}, {Fernique},
  {Fienga}, {Figueras}, {Fouron}, {Fragkoudi}, {Fraile}, {Franke}, {Gai},
  {Garabato}, {Garcia-Gutierrez}, {Garc{\'\i}a-Torres}, {Garofalo}, {Gavras},
  {Gerlach}, {Geyer}, {Giacobbe}, {Gilmore}, {Girona}, {Giuffrida}, {Gomel},
  {Gomez}, {Gonzalez-Santamaria}, {Gonz{\'a}lez-Vidal}, {Granvik},
  {Guti{\'e}rrez-S{\'a}nchez}, {Guy}, {Hauser}, {Haywood}, {Helmi}, {Hidalgo},
  {Hilger}, {H{\l}adczuk}, {Hobbs}, {Holland}, {Huckle}, {Jasniewicz},
  {Jonker}, {Juaristi Campillo}, {Julbe}, {Karbevska}, {Kervella}, {Khanna},
  {Kochoska}, {Kontizas}, {Kordopatis}, {Korn}, {Kostrzewa-Rutkowska},
  {Kruszy{\'n}ska}, {Lambert}, {Lanza}, {Lasne}, {Le Campion}, {Le Fustec},
  {Lebreton}, {Lebzelter}, {Leccia}, {Leclerc}, {Lecoeur-Taibi}, {Liao},
  {Licata}, {Lindstr{\o}m}, {Lister}, {Livanou}, {Lobel}, {Madrero Pardo},
  {Managau}, {Mann}, {Marchant}, {Marconi}, {Marcos Santos}, {Marinoni},
  {Marocco}, {Marshall}, {Martin Polo}, {Mart{\'\i}n-Fleitas}, {Masip},
  {Massari}, {Mastrobuono-Battisti}, {Mazeh}, {McMillan}, {Messina}, {Millar},
  {Mints}, {Molina}, {Molinaro}, {Moln{\'a}r}, {Montegriffo}, {Mor},
  {Morbidelli}, {Morel}, {Morris}, {Mulone}, {Munoz}, {Muraveva}, {Murphy},
  {Musella}, {Noval}, {Ord{\'e}novic}, {Orr{\`u}}, {Osinde}, {Pagani},
  {Pagano}, {Palaversa}, {Palicio}, {Panahi}, {Pawlak}, {Pe{\~n}alosa
  Esteller}, {Penttil{\"a}}, {Piersimoni}, {Pineau}, {Plachy}, {Plum},
  {Poggio}, {Poretti}, {Poujoulet}, {Pr{\v{s}}a}, {Pulone}, {Racero},
  {Ragaini}, {Rainer}, {Raiteri}, {Rambaux}, {Ramos}, {Ramos-Lerate}, {Re
  Fiorentin}, {Regibo}, {Ripepi}, {Riva}, {Rixon}, {Robichon}, {Robin},
  {Roelens}, {Rohrbasser}, {Romero-G{\'o}mez}, {Rowell}, {Royer}, {Rybicki},
  {Sadowski}, {Sagrist{\`a} Sell{\'e}s}, {Salgado}, {Salguero}, {Samaras},
  {Sanchez Gimenez}, {Sanna}, {Santove{\~n}a}, {Sarasso}, {Schultheis},
  {Sciacca}, {Segol}, {Segovia}, {S{\'e}gransan}, {Semeux}, {Shahaf},
  {Siddiqui}, {Siebert}, {Siltala}, {Slezak}, {Solano}, {Solitro}, {Souami},
  {Souchay}, {Spagna}, {Spoto}, {Steele}, {Steidelm{\"u}ller}, {Stephenson},
  {S{\"u}veges}, {Szabados}, {Szegedi-Elek}, {Taris}, {Tauran}, {Taylor},
  {Teixeira}, {Thuillot}, {Tonello}, {Torra}, {Torra}, {Turon}, {Unger},
  {Vaillant}, {van Dillen}, {Vanel}, {Vecchiato}, {Viala}, {Vicente},
  {Voutsinas}, {Weiler}, {Wevers}, {Wyrzykowski}, {Yoldas}, {Yvard}, {Zhao},
  {Zorec}, {Zucker}, {Zurbach}, \& {Zwitter}}]{Gaia2021}
{Gaia Collaboration}, {Smart}, R.~L., {Sarro}, L.~M., {et~al.} 2021, \aap, 649,
  A6, \dodoi{10.1051/0004-6361/202039498}

\bibitem[{{Garufi} {et~al.}(2013){Garufi}, {Quanz}, {Avenhaus}, {Buenzli},
  {Dominik}, {Meru}, {Meyer}, {Pinilla}, {Schmid}, \& {Wolf}}]{Garufi2013}
{Garufi}, A., {Quanz}, S.~P., {Avenhaus}, H., {et~al.} 2013, \aap, 560, A105,
  \dodoi{10.1051/0004-6361/201322429}

\bibitem[{{Garufi} {et~al.}(2014){Garufi}, {Podio}, {Kamp}, {M{\'e}nard},
  {Brittain}, {Eiroa}, {Montesinos}, {Alonso-Mart{\'\i}nez}, {Thi}, \&
  {Woitke}}]{Garufi2014}
{Garufi}, A., {Podio}, L., {Kamp}, I., {et~al.} 2014, \aap, 567, A141,
  \dodoi{10.1051/0004-6361/201321987}

\bibitem[{{Gasman} {et~al.}(2023){Gasman}, {van Dishoeck}, {Grant}, {Temmink},
  {Tabone}, {Henning}, {Kamp}, {G{\"u}del}, {Lagage}, {Perotti}, {Christiaens},
  {Samland}, {Arabhavi}, {Argyriou}, {Abergel}, {Absil}, {Barrado},
  {Boccaletti}, {Bouwman}, {Caratti o Garatti}, {Geers}, {Glauser},
  {Guadarrama}, {Jang}, {Kanwar}, {Lahuis}, {Morales-Calder{\'o}n}, {Mueller},
  {Nehm{\'e}}, {Olofsson}, {Pantin}, {Pawellek}, {Ray}, {Rodgers-Lee},
  {Scheithauer}, {Schreiber}, {Schwarz}, {Vandenbussche}, {Vlasblom}, {Waters},
  {Wright}, {Colina}, {Greve}, \& {{\"O}stlin}}]{Gasman2023}
{Gasman}, D., {van Dishoeck}, E.~F., {Grant}, S.~L., {et~al.} 2023, \aap, 679,
  A117, \dodoi{10.1051/0004-6361/202347005}

\bibitem[{{Gomez Gonzalez} {et~al.}(2017){Gomez Gonzalez}, {Wertz}, {Absil},
  {Christiaens}, {Defr{\`e}re}, {Mawet}, {Milli}, {Absil}, {Van Droogenbroeck},
  {Cantalloube}, {Hinz}, {Skemer}, {Karlsson}, \& {Surdej}}]{GomezGonzalez2017}
{Gomez Gonzalez}, C.~A., {Wertz}, O., {Absil}, O., {et~al.} 2017, \aj, 154, 7,
  \dodoi{10.3847/1538-3881/aa73d7}

\bibitem[{{Gordon} {et~al.}(2022){Gordon}, {Rothman}, {Hargreaves}, {Hashemi},
  {Karlovets}, {Skinner}, {Conway}, {Hill}, {Kochanov}, {Tan}, {Wcis{\l}o},
  {Finenko}, {Nelson}, {Bernath}, {Birk}, {Boudon}, {Campargue}, {Chance},
  {Coustenis}, {Drouin}, {Flaud}, {Gamache}, {Hodges}, {Jacquemart}, {Mlawer},
  {Nikitin}, {Perevalov}, {Rotger}, {Tennyson}, {Toon}, {Tran}, {Tyuterev},
  {Adkins}, {Baker}, {Barbe}, {Can{\`e}}, {Cs{\'a}sz{\'a}r}, {Dudaryonok},
  {Egorov}, {Fleisher}, {Fleurbaey}, {Foltynowicz}, {Furtenbacher}, {Harrison},
  {Hartmann}, {Horneman}, {Huang}, {Karman}, {Karns}, {Kassi}, {Kleiner},
  {Kofman}, {Kwabia-Tchana}, {Lavrentieva}, {Lee}, {Long}, {Lukashevskaya},
  {Lyulin}, {Makhnev}, {Matt}, {Massie}, {Melosso}, {Mikhailenko}, {Mondelain},
  {M{\"u}ller}, {Naumenko}, {Perrin}, {Polyansky}, {Raddaoui}, {Raston},
  {Reed}, {Rey}, {Richard}, {T{\'o}bi{\'a}s}, {Sadiek}, {Schwenke},
  {Starikova}, {Sung}, {Tamassia}, {Tashkun}, {Vander Auwera}, {Vasilenko},
  {Vigasin}, {Villanueva}, {Vispoel}, {Wagner}, {Yachmenev}, \&
  {Yurchenko}}]{Gordon2022}
{Gordon}, I.~E., {Rothman}, L.~S., {Hargreaves}, R.~J., {et~al.} 2022, \jqsrt,
  277, 107949, \dodoi{10.1016/j.jqsrt.2021.107949}

\bibitem[{{Gorti} \& {Hollenbach}(2004)}]{Gorti2004}
{Gorti}, U., \& {Hollenbach}, D. 2004, \apj, 613, 424, \dodoi{10.1086/422406}

\bibitem[{{Grant} {et~al.}(2023){Grant}, {van Dishoeck}, {Tabone}, {Gasman},
  {Henning}, {Kamp}, {G{\"u}del}, {Lagage}, {Bettoni}, {Perotti},
  {Christiaens}, {Samland}, {Arabhavi}, {Argyriou}, {Abergel}, {Absil},
  {Barrado}, {Boccaletti}, {Bouwman}, {o Garatti}, {Geers}, {Glauser},
  {Guadarrama}, {Jang}, {Kanwar}, {Lahuis}, {Morales-Calder{\'o}n}, {Mueller},
  {Nehm{\'e}}, {Olofsson}, {Pantin}, {Pawellek}, {Ray}, {Rodgers-Lee},
  {Scheithauer}, {Schreiber}, {Schwarz}, {Temmink}, {Vandenbussche},
  {Vlasblom}, {Waters}, {Wright}, {Colina}, {Greve}, {Justannont}, \&
  {{\"O}stlin}}]{Grant2023}
{Grant}, S.~L., {van Dishoeck}, E.~F., {Tabone}, B., {et~al.} 2023, \apjl, 947,
  L6, \dodoi{10.3847/2041-8213/acc44b}

\bibitem[{{Gratton} {et~al.}(2019){Gratton}, {Ligi}, {Sissa}, {Desidera},
  {Mesa}, {Bonnefoy}, {Chauvin}, {Cheetham}, {Feldt}, {Lagrange}, {Langlois},
  {Meyer}, {Vigan}, {Boccaletti}, {Janson}, {Lazzoni}, {Zurlo}, {De Boer},
  {Henning}, {D'Orazi}, {Gluck}, {Madec}, {Jaquet}, {Baudoz}, {Fantinel},
  {Pavlov}, \& {Wildi}}]{Gratton2019}
{Gratton}, R., {Ligi}, R., {Sissa}, E., {et~al.} 2019, \aap, 623, A140,
  \dodoi{10.1051/0004-6361/201834760}

\bibitem[{{Gravity Collaboration} {et~al.}(2020){Gravity Collaboration},
  {Garcia Lopez}, {Natta}, {Caratti o Garatti}, {Ray}, {Fedriani},
  {Koutoulaki}, {Klarmann}, {Perraut}, {Sanchez-Bermudez}, {Benisty},
  {Dougados}, {Labadie}, {Brandner}, {Garcia}, {Henning}, {Caselli}, {Duvert},
  {de Zeeuw}, {Grellmann}, {Abuter}, {Amorim}, {Baub{\"o}ck}, {Berger},
  {Bonnet}, {Buron}, {Cl{\'e}net}, {Coud{\'e} Du Foresto}, {de Wit}, {Eckart},
  {Eisenhauer}, {Filho}, {Gao}, {Garcia Dabo}, {Gendron}, {Genzel},
  {Gillessen}, {Habibi}, {Haubois}, {Haussmann}, {Hippler}, {Hubert},
  {Horrobin}, {Jimenez Rosales}, {Jocou}, {Kervella}, {Kolb}, {Lacour}, {Le
  Bouquin}, {L{\'e}na}, {Ott}, {Paumard}, {Perrin}, {Pfuhl}, {Ramirez}, {Rau},
  {Rousset}, {Scheithauer}, {Shangguan}, {Stadler}, {Straub}, {Straubmeier},
  {Sturm}, {van Dishoeck}, {Vincent}, {von Fellenberg}, {Widmann}, {Wieprecht},
  {Wiest}, {Wiezorrek}, {Woillez}, {Yazici}, \& {Zins}}]{GRAVITY2020}
{Gravity Collaboration}, {Garcia Lopez}, R., {Natta}, A., {et~al.} 2020, \nat,
  584, 547, \dodoi{10.1038/s41586-020-2613-1}

\bibitem[{{GRAVITY Collaboration} {et~al.}(2021){GRAVITY Collaboration},
  {Perraut}, {Labadie}, {Bouvier}, {M{\'e}nard}, {Klarmann}, {Dougados},
  {Benisty}, {Berger}, {Bouarour}, {Brandner}, {Caratti O Garatti}, {Caselli},
  {de Zeeuw}, {Garcia-Lopez}, {Henning}, {Sanchez-Bermudez}, {Sousa}, {van
  Dishoeck}, {Al{\'e}cian}, {Amorim}, {Cl{\'e}net}, {Davies}, {Drescher},
  {Duvert}, {Eckart}, {Eisenhauer}, {F{\"o}rster-Schreiber}, {Garcia},
  {Gendron}, {Genzel}, {Gillessen}, {Grellmann}, {Hei{\ss}el}, {Hippler},
  {Horrobin}, {Hubert}, {Jocou}, {Kervella}, {Lacour}, {Lapeyr{\`e}re}, {Le
  Bouquin}, {L{\'e}na}, {Lutz}, {Ott}, {Paumard}, {Perrin}, {Scheithauer},
  {Shangguan}, {Shimizu}, {Stadler}, {Straub}, {Straubmeier}, {Sturm},
  {Tacconi}, {Vincent}, {von Fellenberg}, \& {Widmann}}]{GRAVITY2021}
{GRAVITY Collaboration}, {Perraut}, K., {Labadie}, L., {et~al.} 2021, \aap,
  655, A73, \dodoi{10.1051/0004-6361/202141624}

\bibitem[{{Gravity Collaboration} {et~al.}(2023){Gravity Collaboration},
  {Wojtczak}, {Labadie}, {Perraut}, {Tessore}, {Soulain}, {Ganci}, {Bouvier},
  {Dougados}, {Al{\'e}cian}, {Nowacki}, {Cozzo}, {Brandner}, {Caratti O
  Garatti}, {Garcia}, {Garcia Lopez}, {Sanchez-Bermudez}, {Amorim}, {Benisty},
  {Berger}, {Bourdarot}, {Caselli}, {Cl{\'e}net}, {de Zeeuw}, {Davies},
  {Drescher}, {Duvert}, {Eckart}, {Eisenhauer}, {Eupen},
  {F{\"o}rster-Schreiber}, {Gendron}, {Gillessen}, {Grant}, {Grellmann},
  {Hei{\ss}el}, {Henning}, {Hippler}, {Horrobin}, {Hubert}, {Jocou},
  {Kervella}, {Lacour}, {Lapeyr{\`e}re}, {Le Bouquin}, {L{\'e}na}, {Lutz},
  {Mang}, {Ott}, {Paumard}, {Perrin}, {Scheithauer}, {Shangguan}, {Shimizu},
  {Spezzano}, {Straub}, {Straubmeier}, {Sturm}, {van Dishoeck}, {Vincent}, \&
  {Widmann}}]{GRAVITY2023}
{Gravity Collaboration}, {Wojtczak}, J.~A., {Labadie}, L., {et~al.} 2023, \aap,
  669, A59, \dodoi{10.1051/0004-6361/202244675}

\bibitem[{{Greenwood} {et~al.}(2019){Greenwood}, {Kamp}, {Waters}, {Woitke}, \&
  {Thi}}]{Greenwood2019}
{Greenwood}, A.~J., {Kamp}, I., {Waters}, L.~B.~F.~M., {Woitke}, P., \& {Thi},
  W.~F. 2019, \aap, 626, A6, \dodoi{10.1051/0004-6361/201834365}

\bibitem[{{Hammond} {et~al.}(2023){Hammond}, {Christiaens}, {Price}, {Toci},
  {Pinte}, {Juillard}, \& {Garg}}]{Hammond2023}
{Hammond}, I., {Christiaens}, V., {Price}, D.~J., {et~al.} 2023, \mnras, 522,
  L51, \dodoi{10.1093/mnrasl/slad027}

\bibitem[{{Hoff} {et~al.}(1998){Hoff}, {Henning}, \& {Pfau}}]{Hoff1998}
{Hoff}, W., {Henning}, T., \& {Pfau}, W. 1998, \aap, 336, 242

\bibitem[{{Jakobsen} {et~al.}(2022){Jakobsen}, {Ferruit}, {Alves de Oliveira},
  {Arribas}, {Bagnasco}, {Barho}, {Beck}, {Birkmann}, {B{\"o}ker}, {Bunker},
  {Charlot}, {de Jong}, {de Marchi}, {Ehrenwinkler}, {Falcolini}, {Fels},
  {Franx}, {Franz}, {Funke}, {Giardino}, {Gnata}, {Holota}, {Honnen}, {Jensen},
  {Jentsch}, {Johnson}, {Jollet}, {Karl}, {Kling}, {K{\"o}hler}, {Kolm},
  {Kumari}, {Lander}, {Lemke}, {L{\'o}pez-Caniego}, {L{\"u}tzgendorf},
  {Maiolino}, {Manjavacas}, {Marston}, {Maschmann}, {Maurer}, {Messerschmidt},
  {Moseley}, {Mosner}, {Mott}, {Muzerolle}, {Pirzkal}, {Pittet}, {Plitzke},
  {Posselt}, {Rapp}, {Rauscher}, {Rawle}, {Rix}, {R{\"o}del}, {Rumler},
  {Sabbi}, {Salvignol}, {Schmid}, {Sirianni}, {Smith}, {Strada}, {te Plate},
  {Valenti}, {Wettemann}, {Wiehe}, {Wiesmayer}, {Willott}, {Wright}, {Zeidler},
  \& {Zincke}}]{Jakobsen2022}
{Jakobsen}, P., {Ferruit}, P., {Alves de Oliveira}, C., {et~al.} 2022, \aap,
  661, A80, \dodoi{10.1051/0004-6361/202142663}

\bibitem[{{James} {et~al.}(2022){James}, {Pascucci}, {Liu}, {Banzatti},
  {Krijt}, {Long}, \& {Kamp}}]{James2022}
{James}, M.~M., {Pascucci}, I., {Liu}, Y., {et~al.} 2022, \apj, 941, 187,
  \dodoi{10.3847/1538-4357/ac9c61}

\bibitem[{{Juh{\'a}sz} {et~al.}(2009){Juh{\'a}sz}, {Henning}, {Bouwman},
  {Dullemond}, {Pascucci}, \& {Apai}}]{Juhasz2009}
{Juh{\'a}sz}, A., {Henning}, T., {Bouwman}, J., {et~al.} 2009, \apj, 695, 1024,
  \dodoi{10.1088/0004-637X/695/2/1024}

\bibitem[{{Kamp} {et~al.}(2017){Kamp}, {Thi}, {Woitke}, {Rab}, {Bouma}, \&
  {M{\'e}nard}}]{Kamp2017}
{Kamp}, I., {Thi}, W.-F., {Woitke}, P., {et~al.} 2017, \aap, 607, A41,
  \dodoi{10.1051/0004-6361/201730388}

\bibitem[{{Kamp} {et~al.}(2010){Kamp}, {Tilling}, {Woitke}, {Thi}, \&
  {Hogerheijde}}]{Kamp2010}
{Kamp}, I., {Tilling}, I., {Woitke}, P., {Thi}, W., \& {Hogerheijde}, M. 2010,
  \aap, 510, A260000+, \dodoi{10.1051/0004-6361/200913076}

\bibitem[{{Kamp} {et~al.}(2023){Kamp}, {Henning}, {Arabhavi}, {Bettoni},
  {Christiaens}, {Gasman}, {Grant}, {Morales-Calder{\'o}n}, {Tabone},
  {Abergel}, {Absil}, {Argyriou}, {Barrado}, {Boccaletti}, {Bouwman}, {Caratti
  o Garatti}, {van Dishoeck}, {Geers}, {Glauser}, {G{\"u}del}, {Guadarrama},
  {Jang}, {Kanwar}, {Lagage}, {Lahuis}, {Mueller}, {Nehm{\'e}}, {Olofsson},
  {Pantin}, {Pawellek}, {Perotti}, {Ray}, {Rodgers-Lee}, {Samland},
  {Scheithauer}, {Schreiber}, {Schwarz}, {Temmink}, {Vandenbussche},
  {Vlasblom}, {Waelkens}, {Waters}, \& {Wright}}]{Kamp2023}
{Kamp}, I., {Henning}, T., {Arabhavi}, A.~M., {et~al.} 2023, Faraday
  Discussions, 245, 112, \dodoi{10.1039/D3FD00013C}

\bibitem[{{Kanwar} {et~al.}(2024){Kanwar}, {Kamp}, {Woitke}, {Rab}, {Thi}, \&
  {Min}}]{Kanwar2024}
{Kanwar}, J., {Kamp}, I., {Woitke}, P., {et~al.} 2024, \aap, 681, A22,
  \dodoi{10.1051/0004-6361/202346262}

\bibitem[{{Keppler} {et~al.}(2018){Keppler}, {Benisty}, {M{\"u}ller},
  {Henning}, {van Boekel}, {Cantalloube}, {Ginski}, {van Holstein}, {Maire},
  {Pohl}, {Samland }, {Avenhaus}, {Baudino}, {Boccaletti}, {de Boer},
  {Bonnefoy}, {Chauvin}, {Desidera}, {Langlois}, {Lazzoni}, {Marleau},
  {Mordasini}, {Pawellek}, {Stolker}, {Vigan}, {Zurlo}, {Birnstiel},
  {Brandner}, {Feldt}, {Flock}, {Girard}, {Gratton}, {Hagelberg}, {Isella},
  {Janson}, {Juhasz}, {Kemmer}, {Kral}, {Lagrange}, {Launhardt}, {Matter},
  {M{\'e}nard}, {Milli}, {Molli{\`e}re}, {Olofsson}, {P{\'e}rez}, {Pinilla},
  {Pinte}, {Quanz}, {Schmidt}, {Udry}, {Wahhaj}, {Williams}, {Buenzli},
  {Cudel}, {Dominik}, {Galicher}, {Kasper}, {Lannier}, {Mesa}, {Mouillet},
  {Peretti}, {Perrot}, {Salter}, {Sissa}, {Wildi}, {Abe}, {Antichi},
  {Augereau}, {Baruffolo}, {Baudoz}, {Bazzon}, {Beuzit}, {Blanchard}, {Brems},
  {Buey}, {De Caprio}, {Carbillet}, {Carle}, {Cascone}, {Cheetham}, {Claudi},
  {Costille}, {Delboulb{\'e}}, {Dohlen}, {Fantinel}, {Feautrier}, {Fusco},
  {Giro}, {Gluck}, {Gry}, {Hubin}, {Hugot}, {Jaquet}, {Le Mignant}, {Llored},
  {Madec}, {Magnard}, {Martinez}, {Maurel}, {Meyer}, {M{\"o}ller-Nilsson},
  {Moulin}, {Mugnier}, {Orign{\'e}}, {Pavlov}, {Perret}, {Petit}, {Pragt},
  {Puget}, {Rabou}, {Ramos}, {Rigal}, {Rochat}, {Roelfsema}, {Rousset}, {Roux},
  {Salasnich}, {Sauvage}, {Sevin}, {Soenke}, {Stadler}, {Suarez}, {Turatto}, \&
  {Weber}}]{Keppler2018}
{Keppler}, M., {Benisty}, M., {M{\"u}ller}, A., {et~al.} 2018, \aap, 617, A44,
  \dodoi{10.1051/0004-6361/201832957}

\bibitem[{Koike \& Shibai(1990)}]{Koike1990}
Koike, C., \& Shibai, H. 1990, Monthly Notices of the Royal Astronomical
  Society, 246, 332

\bibitem[{{Koike} {et~al.}(2000){Koike}, {Tsuchiyama}, {Shibai}, {Suto},
  {Tanab{\'e}}, {Chihara}, {Sogawa}, {Mouri}, \& {Okada}}]{Koike2000}
{Koike}, C., {Tsuchiyama}, A., {Shibai}, H., {et~al.} 2000, \aap, 363, 1115

\bibitem[{{K{\'o}sp{\'a}l} {et~al.}(2023){K{\'o}sp{\'a}l}, {{\'A}brah{\'a}m},
  {Diehl}, {Banzatti}, {Bouwman}, {Chen}, {Cruz-S{\'a}enz de Miera}, {Green},
  {Henning}, \& {Rab}}]{Kospal2023}
{K{\'o}sp{\'a}l}, {\'A}., {{\'A}brah{\'a}m}, P., {Diehl}, L., {et~al.} 2023,
  \apjl, 945, L7, \dodoi{10.3847/2041-8213/acb58a}

\bibitem[{{Labiano} {et~al.}(2021){Labiano}, {Argyriou},
  {{\'A}lvarez-M{\'a}rquez}, {Glasse}, {Glauser}, {Patapis}, {Law}, {Brandl},
  {Justtanont}, {Lahuis}, {Mart{\'\i}nez-Galarza}, {Mueller}, {Noriega-Crespo},
  {Royer}, {Shaughnessy}, \& {Vandenbussche}}]{Labiano2021}
{Labiano}, A., {Argyriou}, I., {{\'A}lvarez-M{\'a}rquez}, J., {et~al.} 2021,
  \aap, 656, A57, \dodoi{10.1051/0004-6361/202140614}

\bibitem[{{Lahuis} {et~al.}(2007){Lahuis}, {van Dishoeck}, {Blake}, {Evans},
  {Kessler-Silacci}, \& {Pontoppidan}}]{Lahuis2007}
{Lahuis}, F., {van Dishoeck}, E.~F., {Blake}, G.~A., {et~al.} 2007, \apj, 665,
  492, \dodoi{10.1086/518931}

\bibitem[{{Lisse} {et~al.}(2009){Lisse}, {Chen}, {Wyatt}, {Morlok}, {Song},
  {Bryden}, \& {Sheehan}}]{Lisse2009}
{Lisse}, C.~M., {Chen}, C.~H., {Wyatt}, M.~C., {et~al.} 2009, \apj, 701, 2019,
  \dodoi{10.1088/0004-637X/701/2/2019}

\bibitem[{{Liu} {et~al.}(2019){Liu}, {Pascucci}, \& {Henning}}]{Liu2019}
{Liu}, Y., {Pascucci}, I., \& {Henning}, T. 2019, \aap, 623, A106,
  \dodoi{10.1051/0004-6361/201834418}

\bibitem[{{Luhman}(2023)}]{Luhman2023}
{Luhman}, K.~L. 2023, \aj, 165, 269, \dodoi{10.3847/1538-3881/accf19}

\bibitem[{{Mah} {et~al.}(2023){Mah}, {Bitsch}, {Pascucci}, \&
  {Henning}}]{Mah2023}
{Mah}, J., {Bitsch}, B., {Pascucci}, I., \& {Henning}, T. 2023, \aap, 677, L7,
  \dodoi{10.1051/0004-6361/202347169}

\bibitem[{{Manara} {et~al.}(2023){Manara}, {Ansdell}, {Rosotti}, {Hughes},
  {Armitage}, {Lodato}, \& {Williams}}]{Manara2023}
{Manara}, C.~F., {Ansdell}, M., {Rosotti}, G.~P., {et~al.} 2023, in
  Astronomical Society of the Pacific Conference Series, Vol. 534, Protostars
  and Planets VII, ed. S.~{Inutsuka}, Y.~{Aikawa}, T.~{Muto}, K.~{Tomida}, \&
  M.~{Tamura}, 539, \dodoi{10.48550/arXiv.2203.09930}

\bibitem[{{Mandell} {et~al.}(2012){Mandell}, {Bast}, {van Dishoeck}, {Blake},
  {Salyk}, {Mumma}, \& {Villanueva}}]{Mandell2012}
{Mandell}, A.~M., {Bast}, J., {van Dishoeck}, E.~F., {et~al.} 2012, \apj, 747,
  92, \dodoi{10.1088/0004-637X/747/2/92}

\bibitem[{{Matthews} {et~al.}(2014){Matthews}, {Krivov}, {Wyatt}, {Bryden}, \&
  {Eiroa}}]{Matthews2014}
{Matthews}, B.~C., {Krivov}, A.~V., {Wyatt}, M.~C., {Bryden}, G., \& {Eiroa},
  C. 2014, in Protostars and Planets VI, ed. H.~{Beuther}, R.~S. {Klessen},
  C.~P. {Dullemond}, \& T.~{Henning}, 521--544,
  \dodoi{10.2458/azu_uapress_9780816531240-ch023}

\bibitem[{{McClure} {et~al.}(2020){McClure}, {Dominik}, \&
  {Kama}}]{McClure2020}
{McClure}, M.~K., {Dominik}, C., \& {Kama}, M. 2020, \aap, 642, L15,
  \dodoi{10.1051/0004-6361/202038912}

\bibitem[{{Menu} {et~al.}(2014){Menu}, {van Boekel}, {Henning}, {Chandler},
  {Linz}, {Benisty}, {Lacour}, {Min}, {Waelkens}, {Andrews}, {Calvet},
  {Carpenter}, {Corder}, {Deller}, {Greaves}, {Harris}, {Isella}, {Kwon},
  {Lazio}, {Le Bouquin}, {M{\'e}nard}, {Mundy}, {P{\'e}rez}, {Ricci},
  {Sargent}, {Storm}, {Testi}, \& {Wilner}}]{Menu2014}
{Menu}, J., {van Boekel}, R., {Henning}, T., {et~al.} 2014, \aap, 564, A93,
  \dodoi{10.1051/0004-6361/201322961}

\bibitem[{{Min} {et~al.}(2009){Min}, {Dullemond}, {Dominik}, {de Koter}, \&
  {Hovenier}}]{Min2009}
{Min}, M., {Dullemond}, C.~P., {Dominik}, C., {de Koter}, A., \& {Hovenier},
  J.~W. 2009, \aap, 497, 155, \dodoi{10.1051/0004-6361/200811470}

\bibitem[{{Min} {et~al.}(2005){Min}, {Hovenier}, \& {de Koter}}]{Min2005}
{Min}, M., {Hovenier}, J.~W., \& {de Koter}, A. 2005, \aap, 432, 909,
  \dodoi{10.1051/0004-6361:20041920}

\bibitem[{{Miotello} {et~al.}(2019){Miotello}, {Facchini}, {van Dishoeck},
  {Cazzoletti}, {Testi}, {Williams}, {Ansdell}, {van Terwisga}, \& {van der
  Marel}}]{Miotello2019}
{Miotello}, A., {Facchini}, S., {van Dishoeck}, E.~F., {et~al.} 2019, \aap,
  631, A69, \dodoi{10.1051/0004-6361/201935441}

\bibitem[{{Molli{\`e}re} {et~al.}(2022){Molli{\`e}re}, {Molyarova}, {Bitsch},
  {Henning}, {Schneider}, {Kreidberg}, {Eistrup}, {Burn}, {Nasedkin},
  {Semenov}, {Mordasini}, {Schlecker}, {Schwarz}, {Lacour}, {Nowak}, \&
  {Schulik}}]{Molliere2022}
{Molli{\`e}re}, P., {Molyarova}, T., {Bitsch}, B., {et~al.} 2022, \apj, 934,
  74, \dodoi{10.3847/1538-4357/ac6a56}

\bibitem[{{Mo{\'o}r} {et~al.}(2017){Mo{\'o}r}, {Cur{\'e}}, {K{\'o}sp{\'a}l},
  {{\'A}brah{\'a}m}, {Csengeri}, {Eiroa}, {Gunawan}, {Henning}, {Hughes}, \&
  {Juh{\'a}sz}}]{Moor2017}
{Mo{\'o}r}, A., {Cur{\'e}}, M., {K{\'o}sp{\'a}l}, {\'A}., {et~al.} 2017, \apj,
  849, 123, \dodoi{10.3847/1538-4357/aa8e4e}

\bibitem[{{M{\"u}ller} {et~al.}(2018){M{\"u}ller}, {Keppler}, {Henning},
  {Samland}, {Chauvin}, {Beust}, {Maire}, {Molaverdikhani}, {van Boekel},
  {Benisty}, {Boccaletti}, {Bonnefoy}, {Cantalloube}, {Charnay}, {Baudino},
  {Gennaro}, {Long}, {Cheetham}, {Desidera}, {Feldt}, {Fusco}, {Girard},
  {Gratton}, {Hagelberg}, {Janson}, {Lagrange}, {Langlois}, {Lazzoni}, {Ligi},
  {M{\'e}nard}, {Mesa}, {Meyer}, {Molli{\`e}re}, {Mordasini}, {Moulin},
  {Pavlov}, {Pawellek}, {Quanz}, {Ramos}, {Rouan}, {Sissa}, {Stadler}, {Vigan},
  {Wahhaj}, {Weber}, \& {Zurlo}}]{Mueller2018}
{M{\"u}ller}, A., {Keppler}, M., {Henning}, T., {et~al.} 2018, \aap, 617, L2,
  \dodoi{10.1051/0004-6361/201833584}

\bibitem[{{Muro-Arena} {et~al.}(2018){Muro-Arena}, {Dominik}, {Waters}, {Min},
  {Klarmann}, {Ginski}, {Isella}, {Benisty}, {Pohl}, {Garufi}, {Hagelberg},
  {Langlois}, {Menard}, {Pinte}, {Sezestre}, {van der Plas}, {Villenave},
  {Delboulb{\'e}}, {Magnard}, {M{\"o}ller-Nilsson}, {Pragt}, {Rabou}, \&
  {Roelfsema}}]{Muro-Arena2018}
{Muro-Arena}, G.~A., {Dominik}, C., {Waters}, L.~B.~F.~M., {et~al.} 2018, \aap,
  614, A24, \dodoi{10.1051/0004-6361/201732299}

\bibitem[{{Najita} {et~al.}(2003){Najita}, {Carr}, \& {Mathieu}}]{Najita2003}
{Najita}, J., {Carr}, J.~S., \& {Mathieu}, R.~D. 2003, \apj, 589, 931,
  \dodoi{10.1086/374809}

\bibitem[{{Najita} {et~al.}(2010){Najita}, {Carr}, {Strom}, {Watson},
  {Pascucci}, {Hollenbach}, {Gorti}, \& {Keller}}]{Najita2010}
{Najita}, J.~R., {Carr}, J.~S., {Strom}, S.~E., {et~al.} 2010, \apj, 712, 274,
  \dodoi{10.1088/0004-637X/712/1/274}

\bibitem[{{Nomura} {et~al.}(2007){Nomura}, {Aikawa}, {Tsujimoto}, {Nakagawa},
  \& {Millar}}]{Nomura2007}
{Nomura}, H., {Aikawa}, Y., {Tsujimoto}, M., {Nakagawa}, Y., \& {Millar}, T.~J.
  2007, \apj, 661, 334, \dodoi{10.1086/513419}

\bibitem[{{Oberg} {et~al.}(2023){Oberg}, {Kamp}, {Cazaux}, {Rab}, \&
  {Czoske}}]{Oberg2023}
{Oberg}, N., {Kamp}, I., {Cazaux}, S., {Rab}, C., \& {Czoske}, O. 2023, \aap,
  670, A74, \dodoi{10.1051/0004-6361/202244845}

\bibitem[{{Pascucci} {et~al.}(2009){Pascucci}, {Apai}, {Luhman}, {Henning},
  {Bouwman}, {Meyer}, {Lahuis}, \& {Natta}}]{Pascucci2009}
{Pascucci}, I., {Apai}, D., {Luhman}, K., {et~al.} 2009, \apj, 696, 143,
  \dodoi{10.1088/0004-637X/696/1/143}

\bibitem[{{Pascucci} {et~al.}(2013){Pascucci}, {Herczeg}, {Carr}, \&
  {Bruderer}}]{Pascucci2013}
{Pascucci}, I., {Herczeg}, G., {Carr}, J.~S., \& {Bruderer}, S. 2013, \apj,
  779, 178, \dodoi{10.1088/0004-637X/779/2/178}

\bibitem[{{Pascucci} {et~al.}(2007){Pascucci}, {Hollenbach}, {Najita},
  {Muzerolle}, {Gorti}, {Herczeg}, {Hillenbrand }, {Kim}, {Carpenter}, {Meyer},
  {Mamajek}, \& {Bouwman}}]{Pascucci2007}
{Pascucci}, I., {Hollenbach}, D., {Najita}, J., {et~al.} 2007, \apj, 663, 383,
  \dodoi{10.1086/518535}

\bibitem[{{Pawellek} \& {Krivov}(2015)}]{Pawellek2015}
{Pawellek}, N., \& {Krivov}, A.~V. 2015, \mnras, 454, 3207,
  \dodoi{10.1093/mnras/stv2142}

\bibitem[{{Pawellek} {et~al.}(2024){Pawellek}, {Mo{\'o}r}, {Kirchschlager},
  {Milli}, {K{\'o}sp{\'a}l}, {{\'A}brah{\'a}m}, {Marino}, {Wyatt}, {Rebollido},
  {Hughes}, {Cantalloube}, \& {Henning}}]{Pawellek2024}
{Pawellek}, N., {Mo{\'o}r}, A., {Kirchschlager}, F., {et~al.} 2024, \mnras,
  527, 3559, \dodoi{10.1093/mnras/stad3455}

\bibitem[{Perotti {et~al.}(2023)Perotti, Christiaens, Henning, Tabone, Waters,
  Kamp, Olofsson, Grant, Gasman, Bouwman, Samland, Franceschi, van Dishoeck,
  Schwarz, G{\"u}del, Lagage, Ray, Vandenbussche, Abergel, Absil, Arabhavi,
  Argyriou, Barrado, Boccaletti, Caratti~o Garatti, Geers, Glauser, Justannont,
  Lahuis, Mueller, Nehm{\'e}, Pantin, Scheithauer, Waelkens, Guadarrama, Jang,
  Kanwar, Morales-Calder{\'o}n, Pawellek, Rodgers-Lee, Schreiber, Colina,
  Greve, {\"O}stlin, \& Wright}]{Perotti2023}
Perotti, G., Christiaens, V., Henning, T., {et~al.} 2023, Nature, 620, 516,
  \dodoi{10.1038/s41586-023-06317-9}

\bibitem[{{Pinte} {et~al.}(2023){Pinte}, {Teague}, {Flaherty}, {Hall},
  {Facchini}, \& {Casassus}}]{Pinte2023}
{Pinte}, C., {Teague}, R., {Flaherty}, K., {et~al.} 2023, in Astronomical
  Society of the Pacific Conference Series, Vol. 534, Protostars and Planets
  VII, ed. S.~{Inutsuka}, Y.~{Aikawa}, T.~{Muto}, K.~{Tomida}, \& M.~{Tamura},
  645, \dodoi{10.48550/arXiv.2203.09528}

\bibitem[{{Pinte} {et~al.}(2018){Pinte}, {Price}, {M{\'e}nard}, {Duch{\^e}ne},
  {Dent}, {Hill}, {de Gregorio-Monsalvo}, {Hales}, \& {Mentiplay}}]{Pinte2018b}
{Pinte}, C., {Price}, D.~J., {M{\'e}nard}, F., {et~al.} 2018, \apjl, 860, L13,
  \dodoi{10.3847/2041-8213/aac6dc}

\bibitem[{{Pontoppidan} {et~al.}(2009){Pontoppidan}, {Meijerink}, {Dullemond},
  \& {Blake}}]{Pontoppidan2009}
{Pontoppidan}, K.~M., {Meijerink}, R., {Dullemond}, C.~P., \& {Blake}, G.~A.
  2009, \apj, 704, 1482, \dodoi{10.1088/0004-637X/704/2/1482}

\bibitem[{{Pontoppidan} {et~al.}(2010){Pontoppidan}, {Salyk}, {Blake},
  {Meijerink}, {Carr}, \& {Najita}}]{Pontoppidan2010}
{Pontoppidan}, K.~M., {Salyk}, C., {Blake}, G.~A., {et~al.} 2010, \apj, 720,
  887, \dodoi{10.1088/0004-637X/720/1/887}

\bibitem[{{Pontoppidan} {et~al.}(2023){Pontoppidan}, {Salyk}, {Banzatti},
  {Zhang}, {Pascucci}, {Oberg}, {Long}, {Munoz-Romero}, {Carr}, {Najita},
  {Blake}, {Arulanantham}, {Andrews}, {Ballering}, {Bergin}, {Calahan}, {Cobb},
  {Colmenares}, {Dickson-Vandervelde}, {Dignan}, {Green}, {Heretz}, {Herczeg},
  {Kalyaan}, {Krijt}, {Pauly}, {Pinilla}, {Trapman}, \&
  {Xie}}]{Pontoppidan2023}
{Pontoppidan}, K.~M., {Salyk}, C., {Banzatti}, A., {et~al.} 2023, arXiv
  e-prints, arXiv:2311.17020, \dodoi{10.48550/arXiv.2311.17020}

\bibitem[{{Posch} {et~al.}(2007){Posch}, {Mutschke}, {Trieloff}, \&
  {Henning}}]{Posch2007}
{Posch}, T., {Mutschke}, H., {Trieloff}, M., \& {Henning}, T. 2007, \apj, 656,
  615, \dodoi{10.1086/510445}

\bibitem[{{Qi} {et~al.}(2004){Qi}, {Ho}, {Wilner}, {Takakuwa}, {Hirano},
  {Ohashi}, {Bourke}, {Zhang}, {Blake}, {Hogerheijde}, {Saito}, {Choi}, \&
  {Yang}}]{Qi2004}
{Qi}, C., {Ho}, P. T.~P., {Wilner}, D.~J., {et~al.} 2004, \apjl, 616, L11,
  \dodoi{10.1086/421063}

\bibitem[{{Quanz} {et~al.}(2013){Quanz}, {Amara}, {Meyer}, {Kenworthy},
  {Kasper}, \& {Girard}}]{Quanz2013}
{Quanz}, S.~P., {Amara}, A., {Meyer}, M.~R., {et~al.} 2013, \apjl, 766, L1,
  \dodoi{10.1088/2041-8205/766/1/L1}

\bibitem[{{Rab} {et~al.}(2020){Rab}, {Kamp}, {Dominik}, {Ginski}, {Muro-Arena},
  {Thi}, {Waters}, \& {Woitke}}]{Rab2020}
{Rab}, C., {Kamp}, I., {Dominik}, C., {et~al.} 2020, arXiv e-prints,
  arXiv:2008.05941.
\newblock \doarXiv{2008.05941}

\bibitem[{{Ram{\'\i}rez-Tannus} {et~al.}(2023){Ram{\'\i}rez-Tannus}, {Bik},
  {Cuijpers}, {Waters}, {G{\"o}ppl}, {Henning}, {Kamp}, {Preibisch}, {Getman},
  {Chaparro}, {Cuartas-Restrepo}, {de Koter}, {Feigelson}, {Grant}, {Haworth},
  {Hern{\'a}ndez}, {Kuhn}, {Perotti}, {Povich}, {Reiter}, {Roccatagliata},
  {Sabbi}, {Tabone}, {Winter}, {McLeod}, {van Boekel}, \& {van
  Terwisga}}]{Ramirez-Tannus2023}
{Ram{\'\i}rez-Tannus}, M.~C., {Bik}, A., {Cuijpers}, L., {et~al.} 2023, \apjl,
  958, L30, \dodoi{10.3847/2041-8213/ad03f8}

\bibitem[{{Rich} {et~al.}(2022){Rich}, {Monnier}, {Aarnio}, {Laws},
  {Setterholm}, {Wilner}, {Calvet}, {Harries}, {Miller}, {Davies}, {Adams},
  {Andrews}, {Bae}, {Espaillat}, {Greenbaum}, {Hinkley}, {Kraus}, {Hartmann},
  {Isella}, {McClure}, {Oppenheimer}, {P{\'e}rez}, \& {Zhu}}]{Rich2022}
{Rich}, E.~A., {Monnier}, J.~D., {Aarnio}, A., {et~al.} 2022, \aj, 164, 109,
  \dodoi{10.3847/1538-3881/ac7be4}

\bibitem[{{Rieke} {et~al.}(2005){Rieke}, {Kelly}, {Horner}, \& {NIRCam
  Team}}]{Rieke2005}
{Rieke}, M., {Kelly}, D., {Horner}, S., \& {NIRCam Team}. 2005, in American
  Astronomical Society Meeting Abstracts, Vol. 207, American Astronomical
  Society Meeting Abstracts, 115.09

\bibitem[{{Rigliaco} {et~al.}(2015){Rigliaco}, {Pascucci}, {Duchene},
  {Edwards}, {Ardila}, {Grady}, {Mendigut{\'\i}a}, {Montesinos}, {Mulders},
  {Najita}, {Carpenter}, {Furlan}, {Gorti}, {Meijerink}, \&
  {Meyer}}]{Rigliaco2015}
{Rigliaco}, E., {Pascucci}, I., {Duchene}, G., {et~al.} 2015, \apj, 801, 31,
  \dodoi{10.1088/0004-637X/801/1/31}

\bibitem[{{Riols} \& {Lesur}(2018)}]{Riols2018}
{Riols}, A., \& {Lesur}, G. 2018, \aap, 617, A117,
  \dodoi{10.1051/0004-6361/201833212}

\bibitem[{{Riviere-Marichalar} {et~al.}(2012){Riviere-Marichalar}, {Barrado},
  {Augereau}, {Thi}, {Roberge}, {Eiroa}, {Montesinos}, {Meeus}, {Howard}, \&
  {Sandell}}]{Riviere2012}
{Riviere-Marichalar}, P., {Barrado}, D., {Augereau}, J.~C., {et~al.} 2012,
  \aap, 546, L8, \dodoi{10.1051/0004-6361/201219745}

\bibitem[{{Roberge} {et~al.}(2013){Roberge}, {Kamp}, {Montesinos}, {Dent},
  {Meeus}, {Donaldson}, {Olofsson}, {Mo{\'o}r}, {Augereau}, \&
  {Howard}}]{Roberge2013}
{Roberge}, A., {Kamp}, I., {Montesinos}, B., {et~al.} 2013, \apj, 771, 69,
  \dodoi{10.1088/0004-637X/771/1/69}

\bibitem[{{Ruffio} {et~al.}(2023){Ruffio}, {Perrin}, {Hoch}, {Kammerer},
  {Konopacky}, {Pueyo}, {Rickman}, {Theissen}, {Agrawal}, {Greenbaum}, {Miles},
  {Barman}, {Balmer}, {Llop-Sayson}, {Girard}, {Rebollido}, {Soummer}, {Allen},
  {Anderson}, {Beichman}, {Bellini}, {Bryden}, {Espinoza}, {Glidden}, {Huang},
  {Lewis}, {Libralato}, {Louie}, {Sohn}, {Seager}, {van der Marel}, {Wakeford},
  {Watkins}, {Ygouf}, \& {Mountai}}]{Ruffio2023}
{Ruffio}, J.-B., {Perrin}, M.~D., {Hoch}, K. K.~W., {et~al.} 2023, arXiv
  e-prints, arXiv:2310.09902, \dodoi{10.48550/arXiv.2310.09902}

\bibitem[{{Sacco} {et~al.}(2012){Sacco}, {Flaccomio}, {Pascucci}, {Lahuis},
  {Ercolano}, {Kastner}, {Micela}, {Stelzer}, \& {Sterzik}}]{Sacco2012}
{Sacco}, G.~G., {Flaccomio}, E., {Pascucci}, I., {et~al.} 2012, \apj, 747, 142,
  \dodoi{10.1088/0004-637X/747/2/142}

\bibitem[{{Salyk} {et~al.}(2009){Salyk}, {Blake}, {Boogert}, \&
  {Brown}}]{Salyk2009}
{Salyk}, C., {Blake}, G.~A., {Boogert}, A.~C.~A., \& {Brown}, J.~M. 2009, \apj,
  699, 330, \dodoi{10.1088/0004-637X/699/1/330}

\bibitem[{{Salyk} {et~al.}(2008){Salyk}, {Pontoppidan}, {Blake}, {Lahuis}, {van
  Dishoeck}, \& {Evans}}]{Salyk2008}
{Salyk}, C., {Pontoppidan}, K.~M., {Blake}, G.~A., {et~al.} 2008, \apjl, 676,
  L49, \dodoi{10.1086/586894}

\bibitem[{{Salyk} {et~al.}(2011){Salyk}, {Pontoppidan}, {Blake}, {Najita}, \&
  {Carr}}]{Salyk2011}
{Salyk}, C., {Pontoppidan}, K.~M., {Blake}, G.~A., {Najita}, J.~R., \& {Carr},
  J.~S. 2011, \apj, 731, 130, \dodoi{10.1088/0004-637X/731/2/130}

\bibitem[{{Schr{\"a}pler} \& {Henning}(2004)}]{Schraepler2004}
{Schr{\"a}pler}, R., \& {Henning}, T. 2004, \apj, 614, 960,
  \dodoi{10.1086/423831}

\bibitem[{{Schwarz} {et~al.}(2023){Schwarz}, {Henning}, {Christiaens},
  {Gasman}, {Samland}, {Perotti}, {Jang}, {Grant}, {Tabone},
  {Morales-Calderon}, {Kamp}, {van Dishoeck}, {Gudel}, {Lagage}, {Argyriou},
  {Barrado}, {Garatti}, {Glauser}, {Ray}, {Vandenbussche}, {Waters},
  {Arabhavi}, {Kanwar}, {Olofsson}, {Rodgers-Lee}, {Schreiber}, \&
  {Temmink}}]{Schwarz2023}
{Schwarz}, K.~R., {Henning}, T., {Christiaens}, V., {et~al.} 2023, arXiv
  e-prints, arXiv:2312.07135, \dodoi{10.48550/arXiv.2312.07135}

\bibitem[{{Song} {et~al.}(2015){Song}, {Balakrishnan}, {Walker}, {Stancil},
  {Thi}, {Kamp}, {van der Avoird}, \& {Groenenboom}}]{Song2015}
{Song}, L., {Balakrishnan}, N., {Walker}, K.~M., {et~al.} 2015, \apj, 813, 96,
  \dodoi{10.1088/0004-637X/813/2/96}

\bibitem[{Suto {et~al.}(2006)Suto, Sogawa, Tachibana, Koike, Karoji,
  Tsuchiyama, Chihara, Mizutani, Akedo, Ogiso, Fukui, \& Ohara}]{Suto2006}
Suto, H., Sogawa, H., Tachibana, S., {et~al.} 2006, Monthly Notices of the
  Royal Astronomical Society, 370, 1599

\bibitem[{{Suttle} {et~al.}(2021){Suttle}, {King}, {Schofield}, {Bates}, \&
  {Russell}}]{Suttle2021}
{Suttle}, M.~D., {King}, A.~J., {Schofield}, P.~F., {Bates}, H., \& {Russell},
  S.~S. 2021, \gca, 299, 219, \dodoi{10.1016/j.gca.2021.01.014}

\bibitem[{{Tabone} {et~al.}(2021){Tabone}, {van Hemert}, {van Dishoeck}, \&
  {Black}}]{Tabone2021}
{Tabone}, B., {van Hemert}, M.~C., {van Dishoeck}, E.~F., \& {Black}, J.~H.
  2021, \aap, 650, A192, \dodoi{10.1051/0004-6361/202039549}

\bibitem[{{Tabone} {et~al.}(2023){Tabone}, {Bettoni}, {van Dishoeck},
  {Arabhavi}, {Grant}, {Gasman}, {Henning}, {Kamp}, {G{\"u}del}, {Lagage},
  {Ray}, {Vandenbussche}, {Abergel}, {Absil}, {Argyriou}, {Barrado},
  {Boccaletti}, {Bouwman}, {Caratti o Garatti}, {Geers}, {Glauser},
  {Justannont}, {Lahuis}, {Mueller}, {Nehm{\'e}}, {Olofsson}, {Pantin},
  {Scheithauer}, {Waelkens}, {Waters}, {Black}, {Christiaens}, {Guadarrama},
  {Morales-Calder{\'o}n}, {Jang}, {Kanwar}, {Pawellek}, {Perotti}, {Perrin},
  {Rodgers-Lee}, {Samland}, {Schreiber}, {Schwarz}, {Colina}, {{\"O}stlin}, \&
  {Wright}}]{Tabone2023}
{Tabone}, B., {Bettoni}, G., {van Dishoeck}, E.~F., {et~al.} 2023, Nature
  Astronomy, 7, 805, \dodoi{10.1038/s41550-023-01965-3}

\bibitem[{{Tamura}(2009)}]{Tamura2009}
{Tamura}, M. 2009, in American Institute of Physics Conference Series, Vol.
  1158, Exoplanets and Disks: Their Formation and Diversity, ed. T.~{Usuda},
  M.~{Tamura}, \& M.~{Ishii}, 11--16, \dodoi{10.1063/1.3215811}

\bibitem[{{Teague} {et~al.}(2018){Teague}, {Bae}, {Bergin}, {Birnstiel}, \&
  {Foreman-Mackey}}]{Teague2018}
{Teague}, R., {Bae}, J., {Bergin}, E.~A., {Birnstiel}, T., \& {Foreman-Mackey},
  D. 2018, \apjl, 860, L12, \dodoi{10.3847/2041-8213/aac6d7}

\bibitem[{{Thi} {et~al.}(2013){Thi}, {Kamp}, {Woitke}, {van der Plas},
  {Bertelsen}, \& {Wiesenfeld}}]{Thi2013}
{Thi}, W.~F., {Kamp}, I., {Woitke}, P., {et~al.} 2013, \aap, 551, A49,
  \dodoi{10.1051/0004-6361/201219210}

\bibitem[{{Thi} {et~al.}(2011){Thi}, {M{\'e}nard}, {Meeus}, {Martin-Za{\"i}di},
  {Woitke}, {Tatulli}, {Benisty}, {Kamp}, {Pascucci}, {Pinte}, {Grady},
  {Brittain}, {White}, {Howard}, {Sandell}, \& {Eiroa}}]{Thi2011b}
{Thi}, W.-F., {M{\'e}nard}, F., {Meeus}, G., {et~al.} 2011, \aap, 530, L2,
  \dodoi{10.1051/0004-6361/201116678}

\bibitem[{{Tsukagoshi} {et~al.}(2016){Tsukagoshi}, {Nomura}, {Muto}, {Kawabe},
  {Ishimoto}, {Kanagawa}, {Okuzumi}, {Ida}, {Walsh}, \&
  {Millar}}]{Tsukagoshi2016}
{Tsukagoshi}, T., {Nomura}, H., {Muto}, T., {et~al.} 2016, \apjl, 829, L35,
  \dodoi{10.3847/2041-8205/829/2/L35}

\bibitem[{{van Boekel} {et~al.}(2009){van Boekel}, {G{\"u}del}, {Henning},
  {Lahuis}, \& {Pantin}}]{vanBoekel2009}
{van Boekel}, R., {G{\"u}del}, M., {Henning}, T., {Lahuis}, F., \& {Pantin}, E.
  2009, \aap, 497, 137, \dodoi{10.1051/0004-6361/200811440}

\bibitem[{{van Boekel} {et~al.}(2017){van Boekel}, {Henning}, {Menu}, {de
  Boer}, {Langlois}, {M{\"u}ller}, {Avenhaus}, {Boccaletti}, {Schmid},
  {Thalmann}, {Benisty}, {Dominik}, {Ginski}, {Girard}, {Gisler}, {Lobo Gomes},
  {Menard}, {Min}, {Pavlov}, {Pohl}, {Quanz}, {Rabou}, {Roelfsema}, {Sauvage},
  {Teague}, {Wildi}, \& {Zurlo}}]{vanBoekel2017}
{van Boekel}, R., {Henning}, T., {Menu}, J., {et~al.} 2017, \apj, 837, 132,
  \dodoi{10.3847/1538-4357/aa5d68}

\bibitem[{{van der Marel} {et~al.}(2016){van der Marel}, {Cazzoletti},
  {Pinilla}, \& {Garufi}}]{vanderMarel2016}
{van der Marel}, N., {Cazzoletti}, P., {Pinilla}, P., \& {Garufi}, A. 2016,
  \apj, 832, 178, \dodoi{10.3847/0004-637X/832/2/178}

\bibitem[{{van der Marel} {et~al.}(2013){van der Marel}, {van Dishoeck},
  {Bruderer}, {Birnstiel}, {Pinilla}, {Dullemond}, {van Kempen}, {Schmalzl},
  {Brown}, {Herczeg}, {Mathews}, \& {Geers}}]{vanderMarel2013}
{van der Marel}, N., {van Dishoeck}, E.~F., {Bruderer}, S., {et~al.} 2013,
  Science, 340, 1199, \dodoi{10.1126/science.1236770}

\bibitem[{{van der Plas} {et~al.}(2015){van der Plas}, {van den Ancker},
  {Waters}, \& {Dominik}}]{vanderPlas2015}
{van der Plas}, G., {van den Ancker}, M.~E., {Waters}, L.~B.~F.~M., \&
  {Dominik}, C. 2015, \aap, 574, A75, \dodoi{10.1051/0004-6361/201425052}

\bibitem[{{van der Tak} {et~al.}(2007){van der Tak}, {Black}, {Sch{\"o}ier},
  {Jansen}, \& {van Dishoeck}}]{vanderTak2007}
{van der Tak}, F.~F.~S., {Black}, J.~H., {Sch{\"o}ier}, F.~L., {Jansen}, D.~J.,
  \& {van Dishoeck}, E.~F. 2007, \aap, 468, 627,
  \dodoi{10.1051/0004-6361:20066820}

\bibitem[{{Vlasblom} {et~al.}(2023){Vlasblom}, {van Dishoeck}, {Tabone}, \&
  {Bruderer}}]{Vlasblom2023}
{Vlasblom}, M., {van Dishoeck}, E.~F., {Tabone}, B., \& {Bruderer}, S. 2023,
  arXiv e-prints, arXiv:2311.12445, \dodoi{10.48550/arXiv.2311.12445}

\bibitem[{{Wells} {et~al.}(2015){Wells}, {Pel}, {Glasse}, {Wright},
  {Aitink-Kroes}, {Azzollini}, {Beard}, {Brandl}, {Gallie}, {Geers}, {Glauser},
  {Hastings}, {Henning}, {Jager}, {Justtanont}, {Kruizinga}, {Lahuis}, {Lee},
  {Martinez-Delgado}, {Mart{\'\i}nez-Galarza}, {Meijers}, {Morrison},
  {M{\"u}ller}, {Nakos}, {O'Sullivan}, {Oudenhuysen}, {Parr-Burman}, {Pauwels},
  {Rohloff}, {Schmalzl}, {Sykes}, {Thelen}, {van Dishoeck}, {Vandenbussche},
  {Venema}, {Visser}, {Waters}, \& {Wright}}]{Wells2015}
{Wells}, M., {Pel}, J.~W., {Glasse}, A., {et~al.} 2015, \pasp, 127, 646,
  \dodoi{10.1086/682281}

\bibitem[{{Woitke} {et~al.}(2009){Woitke}, {Kamp}, \& {Thi}}]{Woitke2009}
{Woitke}, P., {Kamp}, I., \& {Thi}, W.-F. 2009, \aap, 501, 383,
  \dodoi{10.1051/0004-6361/200911821}

\bibitem[{{Woitke} {et~al.}(2023){Woitke}, {Thi}, {Arabhavi}, {Kamp}, {Kospal},
  \& {Abraham}}]{Woitke2023}
{Woitke}, P., {Thi}, W.~F., {Arabhavi}, A.~M., {et~al.} 2023, arXiv e-prints,
  arXiv:2311.18321, \dodoi{10.48550/arXiv.2311.18321}

\bibitem[{{Woitke} {et~al.}(2016){Woitke}, {Min}, {Pinte}, {Thi}, {Kamp},
  {Rab}, {Anthonioz}, {Antonellini}, {Baldovin-Saavedra}, {Carmona}, {Dominik},
  {Dionatos}, {Greaves}, {G{\"u}del}, {Ilee}, {Liebhart}, {M{\'e}nard},
  {Rigon}, {Waters}, {Aresu}, {Meijerink}, \& {Spaans}}]{Woitke2016}
{Woitke}, P., {Min}, M., {Pinte}, C., {et~al.} 2016, \aap, 586, A103,
  \dodoi{10.1051/0004-6361/201526538}

\bibitem[{{Woitke} {et~al.}(2018){Woitke}, {Kamp}, {Antonellini}, {Anthonioz},
  {Baldovin-Saveedra}, {Carmona}, {Dionatos}, {Dominik}, {Greaves},
  {G{\"u}del}, {Ilee}, {Liebhardt}, {Menard}, {Min}, {Pinte}, {Rab}, {Rigon},
  {Thi}, {Thureau}, \& {Waters}}]{Woitke2018}
{Woitke}, P., {Kamp}, I., {Antonellini}, S., {et~al.} 2018, arXiv e-prints.
\newblock \doarXiv{1812.02741}

\bibitem[{{Woitke} {et~al.}(2019){Woitke}, {Kamp}, {Antonellini}, {Anthonioz},
  {Baldovin-Saveedra}, {Carmona}, {Dionatos}, {Dominik}, {Greaves},
  {G{\"u}del}, {Ilee}, {Liebhardt}, {Menard}, {Min}, {Pinte}, {Rab}, {Rigon},
  {Thi}, {Thureau}, \& {Waters}}]{Woitke2019}
---. 2019, \pasp, 131, 064301, \dodoi{10.1088/1538-3873/aaf4e5}

\bibitem[{{Wright} {et~al.}(2015){Wright}, {Wright}, {Goodson}, {Rieke},
  {Aitink-Kroes}, {Amiaux}, {Aricha-Yanguas}, {Azzollini}, {Banks},
  {Barrado-Navascues}, {Belenguer-Davila}, {Bloemmart}, {Bouchet}, {Brandl},
  {Colina}, {Detre}, {Diaz-Catala}, {Eccleston}, {Friedman},
  {Garc{\'\i}a-Mar{\'\i}n}, {G{\"u}del}, {Glasse}, {Glauser}, {Greene},
  {Groezinger}, {Grundy}, {Hastings}, {Henning}, {Hofferbert}, {Hunter},
  {Jessen}, {Justtanont}, {Karnik}, {Khorrami}, {Krause}, {Labiano}, {Lagage},
  {Langer}, {Lemke}, {Lim}, {Lorenzo-Alvarez}, {Mazy}, {McGowan}, {Meixner},
  {Morris}, {Morrison}, {M{\"u}ller}, {rgaard-Nielson}, {Olofsson},
  {O'Sullivan}, {Pel}, {Penanen}, {Petach}, {Pye}, {Ray}, {Renotte}, {Renouf},
  {Ressler}, {Samara-Ratna}, {Scheithauer}, {Schneider}, {Shaughnessy},
  {Stevenson}, {Sukhatme}, {Swinyard}, {Sykes}, {Thatcher}, {Tikkanen}, {van
  Dishoeck}, {Waelkens}, {Walker}, {Wells}, \& {Zhender}}]{Wright2015}
{Wright}, G.~S., {Wright}, D., {Goodson}, G.~B., {et~al.} 2015, \pasp, 127,
  595, \dodoi{10.1086/682253}

\bibitem[{{Wright} {et~al.}(2023){Wright}, {Rieke}, {Glasse}, {Ressler},
  {Garc{\'\i}a Mar{\'\i}n}, {Aguilar}, {Alberts}, {{\'A}lvarez-M{\'a}rquez},
  {Argyriou}, {Banks}, {Baudoz}, {Boccaletti}, {Bouchet}, {Bouwman}, {Brandl},
  {Breda}, {Bright}, {Cale}, {Colina}, {Cossou}, {Coulais}, {Cracraft}, {De
  Meester}, {Dicken}, {Engesser}, {Etxaluze}, {Fox}, {Friedman}, {Fu},
  {Gasman}, {G{\'a}sp{\'a}r}, {Gastaud}, {Geers}, {Glauser}, {Gordon},
  {Greene}, {Greve}, {Grundy}, {G{\"u}del}, {Guillard}, {Haderlein},
  {Hashimoto}, {Henning}, {Hines}, {Holler}, {Detre}, {Jahromi}, {James},
  {Jones}, {Justtanont}, {Kavanagh}, {Kendrew}, {Klaassen}, {Krause},
  {Labiano}, {Lagage}, {Lambros}, {Larson}, {Law}, {Lee}, {Libralato}, {Lorenzo
  Alverez}, {Meixner}, {Morrison}, {Mueller}, {Murray}, {Mycroft}, {Myers},
  {Nayak}, {Naylor}, {Nickson}, {Noriega-Crespo}, {{\"O}stlin}, {O'Sullivan},
  {Ottens}, {Patapis}, {Penanen}, {Pietraszkiewicz}, {Ray}, {Regan},
  {Roteliuk}, {Royer}, {Samara-Ratna}, {Samuelson}, {Sargent}, {Scheithauer},
  {Schneider}, {Schreiber}, {Shaughnessy}, {Sheehan}, {Shivaei}, {Sloan},
  {Tamas}, {Teague}, {Temim}, {Tikkanen}, {Tustain}, {van Dishoeck},
  {Vandenbussche}, {Weilert}, {Whitehouse}, \& {Wolff}}]{Wright2023}
{Wright}, G.~S., {Rieke}, G.~H., {Glasse}, A., {et~al.} 2023, \pasp, 135,
  048003, \dodoi{10.1088/1538-3873/acbe66}

\bibitem[{{Wyatt}(2008)}]{Wyatt2008}
{Wyatt}, M.~C. 2008, \araa, 46, 339,
  \dodoi{10.1146/annurev.astro.45.051806.110525}

\bibitem[{{Zannese} {et~al.}(2023){Zannese}, {Tabone}, {Habart}, {Le Petit},
  {van Dishoeck}, \& {Bron}}]{Zannese2023}
{Zannese}, M., {Tabone}, B., {Habart}, E., {et~al.} 2023, \aap, 671, A41,
  \dodoi{10.1051/0004-6361/202244439}

\bibitem[{{Zeidler} {et~al.}(2015){Zeidler}, {Mutschke}, \&
  {Posch}}]{Zeidler2015}
{Zeidler}, S., {Mutschke}, H., \& {Posch}, T. 2015, \apj, 798, 125,
  \dodoi{10.1088/0004-637X/798/2/125}

\bibitem[{{Zhang} {et~al.}(2013){Zhang}, {Pontoppidan}, {Salyk}, \&
  {Blake}}]{Zhang2013}
{Zhang}, K., {Pontoppidan}, K.~M., {Salyk}, C., \& {Blake}, G.~A. 2013, \apj,
  766, 82, \dodoi{10.1088/0004-637X/766/2/82}

\end{thebibliography}
\bibliographystyle{aasjournal}



\end{document}